# Temperature and pressure variability in mid-latitude low atmosphere and stratosphere-ionosphere coupling

## A. L. Morozova[1], P. Ribeiro[1,2], J. J. Blanco[3], and T.V. Barlyaeva[1]

[1]CITEUC, University of Coimbra, Almas de Freire, Sta. Clara, Coimbra, 3040-004, Portugal.
[2] Geophysical and Astronomical Observatory, University of Coimbra, Almas de Freire, Sta. Clara, Coimbra, 3040-004, Portugal.
[3] University of Alcalá, Pza. San Diego, s/n, 28801 Alcalá de Henares, Madrid, Spain.
Corresponding author: Anna Morozova (annamorozovauc@gmail.com)

**Abstract**

This study presents the continuation of our previous analysis of variations of atmospheric and space weather parameters above Iberian Peninsula along two years near the 24[th] solar cycle maximum. In the previous paper (Morozova et al., 2017) we mainly discussed the first mode of principal component analysis of tropospheric and lower stratospheric temperature and pressure fields, which was shown to be correlated with lower stratospheric ozone and anti-correlated with cosmic ray flux. Now we extend the investigation to the second mode, which suggests a coupling between the stratosphere and the ionosphere.

This second mode, located in the low and middle stratosphere (and explaining ~7% of temperature and ~3% of geopotential height variations), showed to be statistically significantly correlated with variations of the middle stratosphere ozone content and anti-correlated with variations of ionospheric total electron content. Similar co-variability of these stratospheric and ionospheric parameters was also obtained with the wavelet cross-coherence analysis.

To investigate the role of atmospheric circulation dynamics and the causal nature of the found correlations, we applied the convergent cross mapping (CCM) analysis to our series. Strong evidence for the stratosphere-ionosphere coupling were obtained for the winter 2012-2013 that is characterized by the easterly QBO phase (quasi-biennial oscillations of the direction of the stratospheric zonal winds) and a strong SSW (sudden stratospheric warming event). Further analysis (for the three-year time interval 2012-2015) hint that SSWs events play main role in emphasizing the stratosphere-ionosphere coupling.

**1 Introduction**

The ionosphere is a partially ionized layer of the Earth's atmosphere located between the upper mesosphere (~ 60 km) and the lower exosphere (~ 900 km), where the UV and XR radiations and energetic particles of solar and cosmic origin are the main ionization sources. Understanding of variations of ionospheric parameters is not only scientifically important but also necessary from the practical and technological points of view. Variations in ionospheric plasma densities change conditions for the radio signal propagation and, consequently, affect the functioning of the satellite based communication, surveillance, and navigation systems (e.g., Kumar and Parkinson, 2017). In this paper we focus on the variations of the ionospheric conditions above the Iberian Peninsula middle latitudinal region.

The ionosphere shows coupling both with the underlying neutral atmosphere and the overlying magnetosphere (Chapman and Bartels, 1951; Kazimirovsky and Kokourov, 1991). In its turn, electric currents running in the ionosphere cause variations of the ground measured geomagnetic field, e.g., well known "solar quiet" (Sq) daily variations (Chapman and Bartels, 1951; Matsushita, 1968; Yamazaki et al., 2016).

The most widely used parameters characterizing ionospheric variable conditions are the total electron content (TEC, i.e. total number of electrons in a column of air of 1 $m^2$ cross section) and the critical frequency of the ionospheric layer F2 (f0F2, i.e. a direct measure of the peak electron density $N_mF2$ provided by vertical incidence ionosondes). The ionospheric parameters have been shown to be influenced by many factors, both external, such as solar irradiance and energy input from the magnetosphere, and internal, e.g., changes in the phase and amplitude of the atmospheric waves and tides (Forbes et al., 2000; Pedatella and Forbes, 2010).

First of all, the ionization level varies with the solar UV flux showing both regular variations on hourly (changes of the insolation during the day), seasonal (Earth's rotation around the Sun) and decadal (e.g. due to solar cycles) time scales, and sporadic changes due to, e.g., solar UV flares (e.g. Maruyama et al., 2009; Rishbeth et al., 2000; Roux et al., 2012).

The ionospheric parameters are also strongly affected by magnetospheric conditions, especially during geomagnetic storms, which cause both increase and decrease of the peak electron density of the ionospheric F2 layer (Cander, 2016; Martyn, 1953; Sato, 1957). In a recent paper Kumar and Parkinson (2017) studied the $N_mF2$ (f0F2) perturbations with respect to the local time at geomagnetic storm onset, season, and the storm intensity. They found that the storm-associated depletions (negative storm effects) and enhancements (positive storm effects) are driven by different but related physical mechanisms, although the depletion mechanism tends to dominate over the enhancement one. The negative storm effects were found to start immediately after geomagnetic storm onset in the nightside high-latitude ionosphere, while the depletions in the dayside high-latitude ionosphere are delayed by a few hours. The equatorward expansion of negative storm effects is found to be regulated by storm intensity (farthest equatorward and deepest during intense storms), season (largest in summer), and time of a day (generally deeper on the nightside). In contrast, positive storm effects typically occur on the dayside mid-latitude and low-latitude ionospheric regions when the storms are in the main phase, regardless of the season. Since, at middle latitudes, TEC usually increases during the initial and main phases of a geomagnetic storm and decreases during the recovery phase (Cander, 2016; Roux et al., 2012), it is expected for TEC to anti-correlate with the geomagnetic Dst index and ground measured horizontal component of geomagnetic field (Roux et al., 2012).

On the other hand, the neutral component (coupled with the ionized one, see e.g., Kazimirovsky et al. (2003) and Leake (2014)) of the upper atmosphere and ionosphere is affected by conditions in the lower atmosphere (stratosphere and even upper troposphere) as different waves and tides propagate upward into the upper atmosphere. These waves and tides travel both between different latitudes and between atmospheric layers (Maruyama et al., 2009; Rishbeth et al., 2000; Yamazaki et al., 2016). Such atmospheric forcing was shown to be responsible for variations of ionospheric parameters, e.g., f0F2 and TEC (Kazimirovsky et al., 2003; Laštovička et al., 2012). Forbes et al. (2000) argue that ~15-20% of the observed ionospheric variability at all latitudes with periods in a range of ~2-30 days under quiet geomagnetic conditions appears to have meteorological origin. The stratosphere–ionosphere coupling was not only observed at different latitudes and during different time intervals but simulated using modern atmospheric and atmosphere-ionosphere models (e.g., Gavrilov et al., 2018; Mendillo et al., 2002; Pedatella, 2016; Pedatella and Liu, 2018).

The hypothesis that the ionosphere can be forced by conditions in the lowest parts of the Earth's atmosphere (stratosphere and upper troposphere) was thoroughly tested during the last decade (e.g., Laštovička, 2006; Laštovička et al., 2006; Liu et al., 2010; Yiğit et al., 2016; see also a review about the ionosphere-stratosphere coupling and the role in it of atmospheric tides and waves by Kazimirovsky et al., 2003 and references therein). Planetary and gravity waves, and atmospheric tides were pointed out as the most probable forcing agents (Ern et al., 2016; Fritts and Alexander, 2003; Laštovička, 2003 and 2006). The amplitudes and periods of such waves/tides change when they propagate toward the upper and less dense atmosphere and interact with the upper atmospheric tides and acoustic waves (Ern et al., 2016; Fritts and Alexander, 2003; Laštovička, 2006; Snively, 2017). This kind of coupling is seen not only in polar regions but also in the middle and even equatorial latitudes (Altadill and Apostolov, 2001 and 2003; Ern et al., 2016; Liu et al., 2010; Pancheva and Mitchell, 2004; see also review by Yiğit et al., 2016 and references therein). Since conditions in the winter stratosphere at middle to high latitudes favor the upward propagation of the atmospheric waves (Fritts and Alexander, 2003), the ionospheric response to the stratospheric forcing is more prominent during the cold months.

One of the interesting phenomena related to the ionosphere-stratosphere coupling through atmospheric waves and tides is the observation of variations of ionospheric parameters during specific events in the polar stratosphere (most often seen in the Northern Hemisphere) named sudden stratospheric warmings (SSW). SSW is defined as a sudden and fast warming of the polar stratosphere accompanied by changes of the strength and direction of the stratospheric zonal wind at 60ºN (Coy and Pawson, 2015). The ionospheric response to SSW is frequently seen at the middle, low and equatorial latitudes in variations of TEC, f0F2 and other ionospheric parameters (Chen et al., 2016; Goncharenko et al., 2013; Jonah et al., 2014; Knížová et al., 2015; Shpynev et al., 2015). This coupling can be observed even at higher altitudes affecting to upper stratosphere and lower mesosphere in the so called high stratospheric warming, HSW (Savenkova et al., 2017).

The stratosphere–ionosphere coupling is usually associated with changes of the phase and amplitude of the atmospheric tides (especially, semidiurnal tides as is shown, e.g., in Pedatella and Forbes, 2010) and also waves with periods of 2-23 days that are strongly amplified in winter atmosphere and during SSW in the vicinity of jet streams, frontal systems and mountain ridges (Altadill and Apostolov, 2001 and 2003; Bramberger et al., 2017; Fritts and Alexander, 2003; Goncharenko et al., 2013; Huang et al., 2018; Knížová et al., 2015; Pancheva and Mitchell, 2004; Phanikumar et al., 2014). The ionospheric response is more prominent during the strong SSW events (Pancheva and Mukhtarov, 2011). Occasionally, this effect is seen in the ionospheric parameters a couple days before the maximum of a SSW event (as is shown for the SSW event in 2008 by Goncharenko and Zhang, 2008), reflecting perturbations that take place in the polar stratosphere and mesosphere during the pre-SSW and SSW periods. Moreover, the response of the ionosphere to a geomagnetic disturbance can be significantly affected by the lower atmosphere during periods of SSW (Pedatella, 2016; Pedatella and Liu, 2018).

Another phenomenon of the lower atmosphere that can affect ionosphere–stratosphere coupling is the quasi-biennial oscillations (QBO) of the direction of the stratospheric zonal winds near the equator. The QBO phase (westerly or easterly, wQBO or eQBO, respectively) affect the propagation conditions for gravity waves in the lower and middle atmosphere (Lu et al., 2008) and, therefore, the polar vortex conditions (e.g., so-called Holton and Tan effect, Holton and Tan 1982). SSWs are more frequent during the eQBO epoch, whereas wQBO periods are associated with stronger and longer living polar vortex, and strong westward zonal winds in the mid-latitudinal stratosphere (Lu et al., 2008). Thus, we can expect dependence of the ionosphere–stratosphere coupling strength on the QBO phase due to, for instance, changes in the amplitude of some atmospheric tides (Yamazaki et al., 2016).

In this paper, we present the continuation of our previous analysis (Morozova et al., 2017) of couplings between the locally measured atmospheric and space weather parameters. In the previous paper we analyzed the first mode of variations of tropospheric and lower stratospheric temperature and pressure fields over a mid-latitudinal region (Iberian Peninsula). This mode was shown to be related to variations of the lower stratosphere ozone content ($O_3$ at 50 hPa level, hereafter $O_{3\,50}$) and the locally measured cosmic ray (CR) flux. Here, we will focus on the second mode that shows co-variability with middle stratosphere ozone content ($O_3$ at 10 hPa level) and geomagnetic and ionospheric parameters, but weak or none co-variability with the CR flux or lower stratosphere ozone ($O_{3\,50}$).

The paper is organized as follows: section 2 contains the descriptions of the analyzed data sets, and section 3 describes the applied mathematical methods. General description of the space weather conditions and the relations between the ionospheric TEC and other space weather parameters are presented in section 4. The atmospheric modes are described in section 5. Results of the analyses of the ionosphere-stratosphere coupling by different statistical methods are presented and discussed in section 6. Finally, section 7 contains main conclusions.

**2 Data**

Most of the data series used in this analysis start on 1 July, 2012 and end on 30 June, 2014 and originally are of daily (UV series) or bi-daily (00:00 and 12:00 UTC, all other series) time resolution. The exceptions are CR series that starts on 7 July, 2012 and series of the ionospheric total electron content (iTEC) and area averaged temperature at the 10 and 50 hPa pressure levels ($T_{10\,hPa}$ and $T_{50\,hPa}$, respectively) that end on 30 June, 2015.

*2.1 Atmospheric data*

The following data series were used to characterize atmospheric conditions in the troposphere and lower and middle stratosphere above the Iberian Peninsula (see also detailed descriptions in Morozova et al., 2017).

Altitudinal profiles of atmospheric parameters from the sounding station at Madrid airport (08221, LEMD, 40.50ºN, 3.58ºW, 633 m asl) are from the Integrated Global Radiosonde Archive (IGRA) database. Each of the observed profiles was re-scaled to the uniform pressure scale from 930 to 30 hPa, $\Delta p = 10$ hPa (91 levels). Two meteorological parameters were analyzed in this paper: the geopotential height of a specific pressure level (*gph*) and the air temperature at this level (*T*). The correspondence between the pressure in hPa and gph in m can be deduced from Fig. 1e of Morozova et al. (2017) or from Fig. S1 in the Supplemented Material. The *T* and *gph* profiles were extended up to the 10 hPa level (resulting in 93 pressure levels total) using the satellite data from the Modern-Era Retrospective Analysis for Research and Applications (MERRA) and the Aqua AIRS Level 3 Daily Standard Physical Retrieval (AIRS+AMSU), AIRX3STD databases. The satellite data are averaged over the area of the Iberian Peninsula variations of the temperature ($T_{10\,hPa}$ and $T_{50\,hPa}$) and gph at the 10 and 50 hPa levels. The time variations of the original *T* and *gph* altitudinal profiles can be found in Fig. 1 of Morozova et al. (2017). As deduced from the temperature profiles, the tropopause is located between ~200–150 and ~50 hPa (approximately between 12 and 20 km) depending on the month. In this paper we consider the region between ~50 and ~30 hPa as the lower stratosphere and the region between ~30 and ~10 hPa as the middle stratosphere.

The QBO phases were defined using the data on the monthly mean equatorial zonal wind components at different stratospheric pressure levels (70-10 hPa) from the Freie Universität Berlin database. In this paper we are going to focus on the relations between the atmospheric and geophysical parameters during three winters: 2012-2013, 2013-2014 and 2014-2015 (hereafter, *eQBO/SSW* winter, *wQBO/noSSW* winter and *eQBO/noSSW* winter, respectively, see also Fig. S2 in the Supplementary Material). The first and third winters are characterized by the easterly QBO phase, however, a strong SSW event was observed only during the first one, in the beginning of January 2013 (January 6-7, see Butler et al., 2017). During the 2014-2015 winter two week SSWs were observed in the polar stratosphere, but there was no significant change of the stratospheric zonal wind at 60N, and the undisrupted polar vortex was observed until the end on winter (end of March – beginning of April 2015), see Manney et al. (2015) and Figure S3 in the Supplementary Material with plots of the zonal averaged temperature and geopotential height anomalies observed in 2012-2015 in the zone 60-90N. These data are from the Global Data Assimilation System of the Climate Prediction Center website (https://www.cpc.ncep.noaa.gov/products/stratosphere/strat-trop/). During the second winter no significant SSW was observed and the QBO was in the westerly phase. Unfortunately, the chosen time interval does not allow a definite separation of the SSW effect from the influence of the QBO phase.

Stratospheric ozone as mole fraction in air measured in the middle stratosphere at 50 and 10 hPa levels ($O_{3\,50}$ and $O_{3\,10}$, respectively) averaged over the area of the Iberian Peninsula, are also from the AIRX3STD data base.

### 2.2 Space weather data

The local ionospheric conditions were characterized by the ionospheric total electron content (*iTEC*) values provided by the Ebro Observatory, Spain (40.8ºN, 0.5ºE, 50 m asl). The instrument currently installed at the Ebro Observatory is the DPS-4D ionospheric sounder and the measured parameter is *f0F2*. The altitude profiles of electron density are calculated from the ionograms, and the integration of these electron profiles up to 1000 km height gives the values of a so called TEC without plasmaspheric contribution or ionospheric TEC. Since *f0F2* is used to calculate *iTEC*, these parameters are highly correlated as can be seen in Fig. S4 in the Supplemented Material.

Two parameters were used to analyze the geomagnetic field variations: the global *Dst* index and the locally measured horizontal component of the geomagnetic field measured by the *Coimbra Magnetic Observatory* (IAGA code COI) located in Coimbra, Portugal (40.22°N, 8.42°W, 99 m asl), hereinafter, *COI H*. *COI H* is in nT and defined as variation relatively to the 25,000 nT level.

To parameterize the variations of the solar UV radiation we used two proxies. The first one is the *Mg II* composite series (Snow et al., 2014), a proxy for the spectral solar irradiance variability in the spectral range from UV to EUV based on the measurements of the emission core of the Mg II doublet (280 nm). The second proxy is the *F10.7* index from the OMNI data base. Recently a number of studies (e.g., Danilov, 2017; Chen et al., 2018; Zhang et al., 2018) showed that the *F10.7* index is not a good proxy for the solar UV flux variations when variations of ionospheric parameters are studied. Therefore, the *Mg II* series which is based on the direct measurements of the UV solar flux was used for analysis, and the *F10.7* index was used only for the *iTEC* regression model.

The cosmic ray (*CR*) flux variations analyzed in this study are from the ground *Castilla-La Mancha Neutron Monitor*, CaLMa (Guadalajara, Spain, 40.63°N, 3.15°W, 708 m a.s.l.) (Medina et al., 2013). This station, with a vertical cut-off rigidity $R_c = 6.95$ GV, gives a direct measurement of the CR arriving to the Iberian Peninsula (see also detailed descriptions in Morozova et al. (2017)).

### 3 Methods

*3.1 Preprocessing and decomposition*

The data sets used in this study were prepared using procedures explained in detail in Morozova et al. (2017). In short, the gaps in the data series were linearly interpolated whenever necessary, and the altitudinal profiles of the atmospheric parameters were rescaled to the uniform pressure scale. The annual cycles were removed from the analyzed series (except the *CR* and *Dst* series) to produce the *noAC* series (see Morozova et al., 2017). The *noAC* series were smoothed using a decomposition procedure named seasonal-trend decomposition based on LOESS (STL). This method, described in detail in Cleveland (1979), Cleveland and Devlin (1988), and Cleveland et al. (1990), allows one to decompose a series into three additive components: a long-term Trend, a Cyclic component with a predefined "period" and a Residual component. The STL procedure can be viewed as a filter that distributes the variations with different periods into three "channels". Those with periods close to the predefined "period" are included in the Cyclic component. The variations with longer periods are filtered into the Trend component, and the rest is regarded as the Residuals. The choice of the "period" values is defined by series' properties and filtering purposes. In this study we used STL to smooth the original series removing both the day-to-day variations (equivalent to running averaging on the window of ~2 days) – the *Smoothed* series, and the short-term variations with characteristic periods shorter than 1-1.5 weeks – the *long-term* series. The last ones were used mostly for visualization purposes (see Fig. 1) and for the principal component analysis (see below), but not for the other types of the analysis due to their strong autocorrelation resulted from the smoothing procedure. Please note the *Smoothed* series from Morozova et al. (2017) correspond to the *long-term* series in this paper.

The modes of the variability of atmospheric parameters (T and gph) were extracted by the principal component analysis (PCA) applied both to the *noAC* and *long-term series*. To extract the coupled variability of the atmospheric parameters a singular value decomposition of the coupled fields (hereafter "cSVD"), an extension of the PCA, was used. Each of the extracted mode is characterized by the pair of a time varying principal component (PC) and a spatially (here, with altitude) varying empirical orthogonal function (EOF). Further, each mode is reconstructed using the corresponding PC and EOF components to obtain time series of $T$ and $gph$ at different pressure levels. The first mode (mode 1) is thoroughly analyzed in Morozova et al. (2017) and briefly described in sec. 5.1. The detailed analysis of the second mode (mode 2) is presented in sec. 5.2 and 6.

### 3.2 Correlation and regression analysis

Similarities between the variations of the analyzed parameters were analyzed using the Pearson correlation coefficients, $r$, that test linear relations between analyzed variables. The significance of the correlation coefficients was estimated using the Monte Carlo approach with artificial series constructed by the "phase randomization procedure" (Ebisuzaki, 1997). The obtained statistical significance ($p\ value$) takes into account the probability of a random series to have the same or higher absolute value of $r$ as in the case of a tested pair of the original series.

### 3.3 Wavelet analysis

The wavelet analysis was used to inspect the evolution of periodicities existing in a data set at different times. The wavelet cross coherence and phase technique was applied to analyze the coherence of a pair of data series, its evolution and the corresponding phase lag between the series. The results are visualized as time-frequency spectra where the powers are represented by different colors (corresponding color map is shown nearby each spectrum). The statistical significance of the computed powers is calculated against the red-noise background. Statistically significant zones of the spectrum (we use the 95% significance level) are contoured by black lines (see Figs. S6-S7 in the Supporting information). An influence of boundary effects is taken into account: one should trust only the results inside the so-called "cone of influence" (bright colored areas).

On the wavelet coherence plots (see Figs. 2 and 5, and Figs. S8-S10 in the Supporting information) the phase relation between the two analyzed data sets – phase lags – are visualized by arrows. If an arrow is directed from left to right then the data sets are in phase, if from right to left they are in anti-phase, if from top to bottom – the first data set leads the second one in quarter of corresponding period. The detailed description of these methods can be found, e.g., in Torrence and Compo (1998) and Maraun and Kurths (2004).

### 3.4 Convergent cross-mapping analysis

While correlation and wavelet cross-coherence analyses are useful tools in detecting similarities in the time-variations of different parameters, nothing can be said with certainty about the causality or direction of the forcing (if any exists) of the analyzed parameters. Other methods are needed to distinguish causality from spurious correlation of parameters characterizing such dynamical systems as atmosphere. One of such power tools is the convergent cross mapping (CCM). This method is based on empirical dynamics (Sugihara et al., 2012 and references therein) and Takens' theorem (Takens, 1981), which states that the essential information of a multidimensional dynamical system is retained in the time series of any single variable of that system (Tsonis et al., 2015). The procedure of the CCM analysis allows to detect if the analyzed parameters belong to the same dynamical system or not and, further, to estimate the strength and direction of the causal link. The CCM methodology is thoroughly described in Sugihara et al. (2012) and Tsonis et al. (2015 and 2018). Here we give only a short summary.

For a pair of analyzed parameters (e.g., X and Y), the causation between the series is analyzed comparing the similarity between the original series and the so-called "shadow manifolds" $M_Y$ and $M_X$ (correspondingly) constructed from lagged coordinates (nonlinear state space reconstruction) of the corresponding (Y and X, respectively) time series. It is also called a "cross mapping of X by using $M_Y$" or X|$M_Y$ and "cross mapping of Y by using $M_X$" or Y|$M_X$. The basic concept of CCM is that a unilateral causation (e.g., X drives Y) results in a possibility to estimate X from Y, but not Y from X (Schiecke et al., 2015). For a bilateral causation with different strengths of the causal link, the quality of the estimations depends on the strength of such link. This quality, or predictive skill, is estimated by a series of correlation coefficients, denoted here as $\rho$(X|$M_Y$) and $\rho$(Y|$M_X$), between the "inputs" (X or Y) and "predictions" ($M_Y$ or $M_X$, correspondingly) for data sets with gradually increasing length (length of library, L). If the skill increases with L, a direct or indirect causal effect of X on Y (or vice versa) can be inferred. Since CCM uses nonlinear state space reconstruction, the causal relations detected by it can be nonlinear too (in contrast to the Pearson correlation discussed in sec. 3.2).

The essential part of CCM is the analysis of the convergence of the $\rho$ series with the increasing L (Sugihara et al., 2012). As was shown in Mønster et al. (2017), the good fit of the converging $\rho$(L) series by an exponential function should be used as an indicator that CCM is applicable to the data set in question, and that its results are reliable.

If only $\rho$(X|$M_Y$) converge and $\rho$(Y|$M_X$) does not (meaning that X can be well reconstructed from Y but not vice versa) then it means that the Y series contains information on X, and X forces variations of Y. When both $\rho$ series converges, it is possible that those parameters are affecting each other more or less equally or they are forced by a third agent (see examples in Sugihara et al., 2012). The statistical significance of the $\rho$ series can be tested using the Monte-Carlo approach and the phase randomization procedure (see sec. 3.2).

CCM was already successfully applied to the analysis of causal relations in biological (Sugihara et al., 2012) and atmospheric (van Nes et al., 2015) systems as well as to test the *CR* and climate relations (Tsonis et al., 2015). Here we applied the CCM analysis (using the R implementation by Ye et al., https://cran.r-project.org/package=rEDM) to test the causal nature of relations between the atmospheric and space weather parameters detected by the correlation analyses (see sec. 5).

Please note that only non-autocorrelated series can be used as the input data sets for CCM (Tsonis et al., 2015). Therefore, only *noAC* series were submitted to the CCM analysis. Also, the first time derivative of the *Mg II noAC* series was used in the CCM analyses instead of the original series (see discussion in Tsonis et al., 2018 and references therein).

**4 TEC and space weather parameters**

Since the main goal of this work is to analyze the ionospheric variations associated with the stratosphere forcing, the *iTEC* variations related to the space weather influence (e.g., due to variations of the geomagnetic field and the solar UV flux) have to be taken into account first.

The *Smoothed* and *long-term iTEC, COI H* and *Dst*, and *F10.7* and *Mg II* series are shown in Figs. 1a, 1b and 1c, respectively. The *iTEC* series, as expected, statistically significantly correlates with UV and geomagnetic series (Table 1) both for the whole length of the series and for the cold seasons only. As a rule, the correlation coefficients with the geomagnetic series compared to the UV series are lower in the absolute values.

The *Mg II* and *iTEC* series show periodic variations close to the period of solar rotation (~27 days) overlaid by short-term variations (corresponding spectrum can be found in Fig. 2b).

The wavelet cross-coherence analysis shows that *iTEC* variations are in phase with variations of *Mg II* with periods close to 27 day-long solar rotation period and its 2[nd] and 3[rd] harmonics (see Fig. 2). These variations are clearly seen in Fig. 3 were variations of the ionospheric *iTEC* during two winters are shown: *eQBO/SSW* (Fig. 3a) and *wQBO/noSSW* (Fig. 3b). The variations of its main external forcings – solar UV flux (*Mg II*) and geomagnetic field (*COI H*) are also shown. As one can see, the solar UV changes, mostly with ~27 days periodicity, affect the average level of *iTEC*.

The relations between the *iTEC* and *COI H* series are more complex. The wavelet analysis of the *noAC COI H* and *iTEC* series show persistent variations with periods of ~2-4 months and transient variations with periods ~1-4 weeks (Figs. 2e). The short-term variations of *iTEC* during winter time intervals (Fig. 3) are in a good agreement with changes of the *COI H* – these two series anti-correlate (please note reversed Y-axes for *COI H* in Fig. 3) but the values of the statistical significance of the correlation coefficients are low (see Table 1). This relation is much stronger during the *eQBO/SSW* winter. Same results are obtained in the wavelet cross-coherence analysis (in Fig. 2e): during this winter (white rectangle "1") there is a strong coherent signal with periodicities from 4 to 16 days. During the *wQBO/noSSW* winter (white rectangle "2" in Fig. 2e) coherent signals at similar period still exist but they are statistically significant only during short time intervals. Please note that the daily variations were removed from both the *iTEC* and *COI H* bi-daily series during the pre-processing.

The CCM analysis of the causal links between *iTEC* and *Mg II* done for two winter seasons shows statistically significant influence of the *Mg II* variations on *iTEC* during the *eQBO/SSW* winter (Fig. 2a). For the *wQBO/noSSW* winter the prediction skill $\rho(Mg\ II|\mathrm{M}_{TEC})$ is higher than $\rho(iTEC|\mathrm{M}_{Mg\ II})$, in addition, the correlation coefficient between the $\rho(Mg\ II|\mathrm{M}_{TEC})$ series and its exponential fit ($r = 0.71$) is higher than corresponding correlation coefficient for $\rho(iTEC|\mathrm{M}_{Mg\ II})$ ($r = 0.45$) – Fig. 2c. Therefore, we can conclude that *Mg II* still influences the *iTEC* variations; however, the statistical significance of this link is lower than 95%. These findings can be explained by the effect of other forcings: e.g., geomagnetic storms and/or upper atmosphere dynamics/composition. It is possible that the QBO phase also plays its role in the changes of the correlation's significance.

The CCM analysis of the causal links between *iTEC* and *COI H* shows that these parameters are coupled and/or under effect of an external forcing: $\rho(COI\ H\ |\mathrm{M}_{TEC})$ and $\rho(iTEC|\mathrm{M}_{COI\ H})$ are of the same amplitude and are well fitted by the exponent ($0.88 \leq r \leq 0.99$) – Fig. 2d and 2e. These ambiguous results are probably due to the mutual influence of the ionospheric electric and the geomagnetic field. Geomagnetic disturbances affect ionosphere while the ionospheric daily currents and irregular disturbances produce magnetic field variations measured at the ground level. It is possible that the *COI H* and *iTEC* series need specific pre-processing to disentangle common variabilities (e.g., separation to the quiet and disturbed components or filtering of signals with specific periods) for a successful CCM analysis. Since the main goal of this work is the analysis of the coupling between the ionosphere, magnetosphere, solar activity and stratosphere on a relatively long time scale (weeks to months), we did not perform a detailed analysis of the *iTEC* variations during individual geomagnetic storms and rather paid attention to the periods of relatively long decreases of the variations of the *Dst* and *COI H* series (as is seen in Fig. 1b) which correspond to time intervals with frequent and strong storms. Also, the time resolution of the analyzed series does not allow for the analysis of a short-living ionospheric disturbances like traveling atmospheric/ionospheric disturbances (TAD/TID).

## 5 Atmospheric modes
### 5.1 Mode 1
The first mode of the $T$ and $gph$ variations is thoroughly described in Morozova et al. (2017). It is defined as PC1/EOF1 obtained both in the PCA and cSVD analyses. It explains a significant part of the variability of the parental series (67–79%). Here we give only a brief description of the found relations with space weather parameters. The first mode of the regional (Iberian Peninsula) atmospheric variability is related to the hemispheric-scale circulation forced by the polar vortex conditions and SSW events. The $T$ and $gph$ variations associated to this mode correlate with the lower stratosphere ozone $O_{3\,50}$ and anti-correlate with the $CR$ flux variations. It was also found that the strength of these correlations depends on the QBO phase (and/or existence or absence of SSW) and, as a consequence, on the blocking or strengthening of the meridional circulation in the Northern Hemisphere stratosphere. We proposed two mechanisms that can explain the found co-variability. The first one is based on the effect of $CR$ particles on the composition of the upper and middle atmosphere ($NO_x$ and $HO_x$ species) and, consequently, on the ozone content in the polar regions and on the polar vortex conditions that, through the coupling between the troposphere and stratosphere in the middle and high latitudes, may affect the atmosphere even at ~40°N. Another reason for the co-variability of the atmospheric parameters and $CR$ is the so-called atmospheric effect (dependence of the ground-measured neutron monitor data on the atmospheric temperature and pressure) that is not fully accounted for by the standard procedure of pressure correction. This assumption is based on the fact that the highest correlation coefficients between the $CR$ and the $T$ and $gph$ series were obtained for the altitudes of ~100–200 hPa ( ~12–16 km), the region where most of the secondary neutrons are produced. Unfortunately, the time scale of the analyzed variations (weeks to months) does not allow discriminating between these two mechanisms.

Since the mode 1 of the $T$ and $gph$ variations is analyzed in Morozova et al. (2017), in this paper the *Smoothed T* and *gph* series are shown with the mode 1 subtracted (Figs. 1e and 1f, respectively). These plots can be compared with Figs. 2b and 2d in Morozova et al. (2017), respectively.

### 5.2 Mode 2
In this paper we present the analysis of the variation of the second atmospheric mode. The numbers of respective PCs/EOFs as well as corresponding variance fractions are in Table 2. Figure 4 shows reconstructed variations of the $T$ and $gph$ mode 2 as color time-altitude plots and the corresponding PCs (for the *Smoothed* series) as lines with symbols (Figs. 4a and 4c). These plots can be compared with Figs. 5a and 5c, respectively, in Morozova et al. (2017) showing variations related to the mode 1. The PCs obtained for the $T$ series are shown as well in Fig. 3 (two winter seasons only).

The mode 2 is located in the lower and middle stratosphere. The highest amplitudes related to this mode are seen above ~70 hPa level for the $T$ variations and above ~30 hPa for $gph$, as is seen in Figs. 4b and 4d. More details on the altitude profile of the EOFs for modes 1 and 2 can be deduced from Fig. S5 in the Supplementary material. The mode 2 is essentially a winter mode: its PCs have highest amplitudes between November-December and April, as shown in Figs. 4a and 4c.

Wavelet analysis of the mode 2 *noAC* PCs (see Supplemented Material, Fig. S7) shows statistically significant variations with period ~4-6 months and a more short-living variability with periods ~1-5 weeks, mostly during winter periods and especially during the *eQBO/SSW* winter. When compared to space weather and ozone parameters the mode 2 shows statistically significant correlations with *iTEC*, as shown in Table 3. Contrary to the mode 1, the mode 2 shows no correlations with *Dst* (a global parameter, comparing to the regional *COI H*, characterizing the geomagnetic field variations) and *CR*.

We presume that at least part of the variations of this atmospheric mode, especially during the winter season with lower level of insolation, may result from the stratosphere-ionosphere coupling. This hypothesis is tested using different mathematical approaches in sec. 6.

**6 Stratosphere-ionosphere coupling as seen by different methods**

Figure 3 shows winter variations of the temperature mode 2 (both from the PCA and cSVD analyses) together with variations of *iTEC* and its main external forcings – solar UV flux (*Mg II*) and geomagnetic field (*COI H*).While changes of the solar UV (mostly, a ~27-day cycle) affect the mean level of *iTEC*, the short-term variations of *iTEC* are very well explained by the *COI H* variations during the *eQBO/SSW* winter (Fig. 3a) and less well during the *wQBO/noSSW* winter. Nevertheless, the sharp increase of *iTEC* between December 26, 2012 and January 14, 2013 (marked in Fig. 3a by the dashed line rectangle) is hardly explained by geomagnetic field variations, and the growing trend of UV solar flux is expected to cause only a global increase of the *iTEC* level. This time interval corresponds to the pre-SSW and SSW conditions related to the SSW event started on January 6, 2013. Following Goncharenko et al. (2013) and Chen et al. (2016), we can attribute these changes in *iTEC* observed before and during the SSW event to the stratosphere-ionosphere coupling.

The stratospheric conditions do not affect the ionospheric *iTEC* directly, but, most probably, through a change in the conditions for the gravity waves/tides upward propagations. When such waves/tides reach the mesosphere and thermosphere, they interact with waves and tides in the upper atmosphere changing conditions in the neutral and, as a consequence, ionized components of the ionosphere (Laštovička et al., 2012). Since the conditions for propagation of these waves and tides depends on the QBO phase and/or existence of the SSW events, the comparison of the results obtained for the *eQBO/SSW* and *wQBO/noSSW* winters can allow us to deduce the existence and significance of the stratosphere-ionosphere coupling during the analyzed time interval over the Iberian Peninsula. In sec. 6.1-6.3 we present the analysis of the winter season variations of the ionospheric and stratospheric parameters and confirmations of their couplings obtained by the correlation, wavelet cross-coherence and CCM analyses.

We must mention that while *T* and *gph* mode 2 PCs are very similar (Fig. 4) and coupled (especially the cSVD PCs), there is no strict similarity of the performances of the *T* and *gph* mode 2 PCs in the correlation, wavelet and CCM analyses. The explanation is probably in the spatial distribution of the *T* and *gph* mode 2: while the *T* mode 2 has maximum around 50-20 hPa, the *gph* mode 2 reaches the highest amplitude above 30 hPa level with a see-saw like pattern (variations at 150-30 hPa oppose one at 30-10 hPa level, Fig. 4d, see also Fig. S5).

Additionally, we tried to remove the geomagnetic and UV variations from the *iTEC* series using multiple linear regression model, but the results of the comparison of the *iTEC* residual series with atmospheric series were only slightly better than for the whole *iTEC* series. Thus, in this paper we show only results for the whole *iTEC* series.

*6.1 Correlation analysis*

Since the stratospheric ozone is one of the agents responsible for the variations of the stratospheric temperature, we compared the variations of the $O_{3\ 10}$ to the mode 2. First of all, the sign of correlation coefficients between the PCs of $T/gph$ and the $O_{3\ 10}$ series is the same for both winters, but the highest correlation coefficients are obtained for the *wQBO/noSSW* winter (not shown here). Same can be said about the relations between *iTEC* and *Mg II* (Table 1) which is expected since solar-magnetosphere-ionosphere interactions do not depend on the atmospheric circulation.

Change of the sign of the correlation coefficients calculated for the *eQBO/SSW* and *wQBO/noSSW* winters was found for the PCs of $T/gph$ vs *iTEC* (Table 3). For the *eQBO/SSW* winter the mode 2 tends to correlate with *iTEC*, whereas for the *wQBO/noSSW* winter there is a strong anti-correlation. These differences between the *eQBO/SSW* and *wQBO/noSSW* winters can be partly explained by the increased solar and geomagnetic activity during the second winter season (corresponding to the higher second solar activity peak during the solar maximum): higher values of the UV flux, more flares, CMEs and geomagnetic storms can be deduced from Figs. 1b and 1c (see also Table S1 in Morozova et al., 2017). Nonetheless, it seems that the intra-atmospheric relations (e.g., between temperature and pressure fields and *iTEC*) are controlled by the atmospheric dynamics and conditions for the waves/tides propagation that are both QBO-dependent and are affected by the appearance/absence of the SSW event. Therefore, it is possible to attribute the change of the sign of the correlation between the mode 2 and *iTEC* to the influence of QBO and/or SSW that are both known to affect the whole Northern hemisphere and conditions for the waves/tides propagations (see sec. 1).

*6.2 Wavelet cross-coherence analysis*

The wavelet cross-coherence analysis of the variations of the mode 2 *noAC* PCs vs *iTEC* (see Figs. 5b and 5e for PCA PCs and Figs. S8 in the Supplemented Materialfor cSVD PCs) series confirms results obtained by the correlation analysis for the *eQBO/SSW* and *wQBO/noSSW* winters. In particular, there is a clear inter-winter difference in the cross-coherence spectra for the mode 2 vs the *iTEC* series. During the *eQBO/SSW* winter (white rectangles "1") there are (almost) in-phase variations between the mode 2 PCs and *iTEC* with periods of ~8-16 days (compare Figs. 5b and Figs. S8a to Table 3). On the contrary, during the *wQBO/noSSW* winter (white rectangles "2") there are anti-phase variations between the mode 2 PCs and *iTEC* series at periods of ~3-4 months (compare Figs. 5e and Figs. S8b to Table 3).

The analysis of the relations between the mode 2 *noAC* PCs and $O_{3\ 10}$ *noAC* series (Figs. S9b, S9e and S10 in the Supplemented Material) shows that all mode 2 PCs are in-phase with variations of the $O_{3\ 10}$ series at periods longer than 4 months throughout one year and at periods ~3-4 weeks during winter seasons.

*6.3 Convergent cross mapping*

Correlation and wavelet analyses show similarities between the PCs for the $T/gph$ mode 2 and the *iTEC* and $O_{3\ 10}$ series. To study the possible causal relations between these parameters the CCM method was used. Results of the CCM analysis are shown in Fig. 5 for the mode 2 PCs vs *iTEC* and in the Supplemented Material Fig. S9 for the mode 2 PCs vs $O_{3\ 10}$. They represent changes of the prediction skill $\rho$ (correlation coefficients between the original series and the CCM reconstruction) with the increase of the library length L shown as a percent of the length of a data set submitted to the CCM analysis. Color-shaded areas of the corresponding color show 95% significance level obtained using the Monte-Carlo simulations.

The CCM analysis for the two winter seasons (Figs. 5a vs 5c and Figs. 5d vs 5f) shows that the $T$ and $gph$ series are well reconstructed from the $iTEC$ series for the $eQBO/SSW$ winter: the $\rho(T|M_{TEC})$ and $\rho(gph|M_{TEC})$ converge and have statistical significance > 95% (Figs. 5a and 5d); they are also well fitted by the exponential function ($0.97 \le r \le 0.99$). On the contrary, for the $wQBO/noSSW$ winter the CCM analysis shows no causal link between variations of the mode 2 and $iTEC$ except for the PCA PC for the $gph$ series (Fig. 5f): in this particular case $\rho(gph|M_{TEC})$ is significantly higher than $\rho(iTEC|M_{gph})$ and well fitted by the exponent ($r = 0.94$), but the statistical significance of this result is low.

The CCM analysis of the relations between the $T$ mode 2 and the ozone series shows that $T/gph$ and $O_{3\ 10}$ affect each other: both $\rho(T|O_{3\ 10})$ and $\rho(O_{3\ 10}|T)$ series converge. This is to be expected since the ozone heats the stratosphere absorbing solar UV light, but an increase of the temperature in the presence of some atmospheric constituents (like $NO_x$ species) can result in an intensification of the ozone destruction (Flury et al., 2009). Besides, both temperature and ozone content are affected by the air masses circulation which, in turn, is affected by the QBO phase, season and features of the global atmospheric dynamics. For example, at the latitudes of the Iberian Peninsula the proportion of the ozone originated at higher/lower latitudes changes with season (Grewe, 2006). Also, the position of the peninsula (between the polar and subtropical jets, see Mohanakumar (2008), results in an advection of the polar ozone poor air masses during SSW (Keil et al., 2007) that can be seen as well in Fig. 4c. Previous studies (Randel and Cobb, 1994; Lee and Smith, 2003; Mohanakumar, 2008) also showed the QBO-dependant ozone variations during winter months.

Still, the results of the CCM analysis suggest that during the studied winter seasons the influence of the stratospheric temperature and pressure (atmosphere dynamics) on the ozone content is stronger than the influence of the ozone heating on the stratospheric conditions (see Figs. S9a vs S9c and Figs. S9d vs S9f in the Supplemented Material). Also, the results of the CCM analysis for the $gph$ series have, generally, higher statistical significance. We also note that for the $wQBO/noSSW$ winter (without strong perturbations in the Northern Hemisphere atmospheric dynamics associated with SSW) the link between the temperature and the ozone content detected by CCM is more straightforward and statistically significant ($\ge$95%, see Fig. S9c and S9f). The skill $\rho(T|M_{O3\ 10})$ is higher and is better fit by an exponential function ($r = 0.99$) than $\rho(O_{3\ 10}|M_T)$ ($r = \sim0.2$). For the $eQBO/SSW$ winter (Supplemented Material, Fig. S9a and S9d) the influence of the stratospheric temperature on the ozone seems to be weaker.

Trying to separate the QBO and SSW influence on the stratosphere-ionosphere coupling we compared the variations of $iTEC$ to the changes of the area averaged stratospheric temperatures at the 10 and 50 hPa pressure levels ($T_{10\ hPa}$ and $T_{50\ hPa}$) available for a longer time interval (from 1 July, 2012 to 30 June, 2015). This time interval covers three winters with the third one being eQBO winter without major SSW– $eQBO/noSSW$ winter (using the conventional definition of major SSW, see also Manney et al., 2015). As discussed in Manney at al. (2015), during this winter there were two events with suddenly rising stratospheric temperature inside the polar vortex. However, neither of these events resulted in a major SSW affecting the whole Northern Hemisphere, and the vortex stayed intact until the end of the winter. The variations of the polar temperature and gph can be seen in plot done by the Global Data Assimilation System of the Climate Prediction Center that can be found also in the Supplementary Material for 2012-2015 epoch (Fig. S3). There are significant differences in the behavior of the stratospheric temperature and gph in the polar air during the winter 2012-2013 and the two following winters, 2013-2014 and 2014-2015. If the SSW is a main driver of the stratosphere-ionosphere coupling, then we should expect that relations between the stratospheric and ionospheric parameters will be similar for the second and the third winters, and differ to ones seen during the first winter. On the contrary, if the QBO phase is the main driver for the stratosphere-ionosphere relations, then there will be similarities between the first and the third winters, which, in turn, will be different to the second one.

The variations of the $T_{10\,hPa}$ and $T_{50\,hPa}$ *Smoothed* series from 1 July, 2012 to 30 June, 2015 are shown in Fig. 6. As one can see, variations of the temperature mode 2 and stratospheric temperatures at 10 and 50 hPa are similar but not equal. The correlation coefficients are 0.5-0.6 (*p value* < 0.01) for the whole series and 0.4-0.6 (*p value* ≤ 0.02) for the first (*eQBO/SSW*) winter and 0.7-0.8 (*p value* < 0.01) for the second (*wQBO/noSSW*) winter. Thus, $T_{10\,hPa}$ and $T_{50\,hPa}$ can be considered only as a proxy for the variations of the mode 2 for the third (*eQBO/noSSW*) winter.

The correlation coefficients between the variations of the *iTEC* and $T_{10\,hPa}$ and $T_{50\,hPa}$ series for three winter time intervals are shown in Table 3 and the series can be compared in Fig. 6. As one can see, the correlation coefficients are similar for the noSSW winters: negative and *p value* > 0.2, whereas for the SSW winter *r* is positive and *p value* ≤ 0.1. The CCM analysis of the *iTEC* and $T_{10\,hPa}$ (middle stratosphere) series (Fig. 7a) shows that for the *eQBO/SSW* winter there is statistically significant atmospheric forcing of the ionospheric parameter (similar to what was obtained for the mode 2, compare with Fig. 5a). However, for the other two winters (*wQBO/noSSW* and *eQBO/noSSW*) no statistically significant forcing can be detected (Fig. 7b and 7c). For the $T_{50\,hPa}$ (lower stratosphere) series no statistically significant results were obtained.

Thus, the results obtained with the stratospheric temperature series hint that the stratosphere-ionosphere coupling at the analyzed region (Iberian Peninsula) is more affected by the major SSW than by the QBO phase.

### 6.4 Discussion

The results of three different methods and, in particular, the CCM analysis confirm the hypothesis that the changes of the stratospheric *T* and *gph* force variations of the ionospheric *iTEC* during the *eQBO/SSW* winter. This forcing could be deduced for the *gph* mode 2 for the *wQBO/noSSW* winter as well, but the statistical significance of this result is < 95%. One must keep in mind that both the stratospheric temperature and ionospheric *iTEC* can be affected by the solar UV flares: the former indirectly through the variations of the ozone content and the latter directly through the ionization by the UV light. All this could influence the CCM estimations of the strength and direction of the causal link between the *T* and *iTEC* series.

We assumed that the correlation of the *T/gph* mode 2 and the *iTEC* series is due to the stratosphere-ionosphere coupling, probably through the gravity waves and tides. If this assumption is correct, then the lower statistical significance of the CCM results for the *wQBO/noSSW* winter can be associated with a lower intensity of the gravity waves during the second winter, as was shown previously in, e.g., Chernigovskaya et al. (2015), Yiğit et al. (2016) and Solomonov et al. (2017). We found confirmation for this hypothesis in the recently published data by Ern et al. (2018). As one can see in their Figs. 19-20, for the 40º N latitudinal zone the winter 2012-2013 (*eQBO/SSW*) is characterized by an intense and long-lasting (until spring 2013) activity of gravity waves at all analyzed altitudes (from 30 to 70 km). On the contrary, the gravity wave activity observed during the winter 2013-2014 (*wQBO/noSSW*) is restricted to the winter months.

The enhancement of variations with periods 8-18 days observed during the *eQBO/SSW* winter in the variations of the ozone, *iTEC* and *T/gph* series (Figs. 5b, and Figs. S6c, S8, S9b and S10b in the Supplemented Material) can be considered as another confirmation for the proposed hypothesis. Previous studies (e.g., Kazimirovsky, 2003; Sridharan, 2017) found that quasi-periodic oscillations in the ionospheric parameters with periods of 6-16 days may be connected with planetary wave activity in the lower atmosphere. Moreover, (Sridharan, 2017) showed that during the SSW 2013 event the planetary waves with periods close to 16 days propagate from high-latitudes toward the equator and interact with the semidiurnal tides there. The time lag between the high and low latitudes is of the order of 3 weeks (see Sridharan, 2017). This time lag is in agreement with time of appearance in our mid-latitudinal region of the coherent signal in the cross-coherent spectra of *T* mode 2 vs *iTEC*: about 1-1.5 weeks after the SSW onset (see Figs. 5b, 5e and Supplemented Material, Fig. S8). Also, Pancheva et al.(2003), Goncharenko et al. (2012) and Jin et al. (2012) showed that variations with periods 8-12 and 15-18 days are observed during SSW winters in the amplitude modulation of the semidiurnal tide, which is considered to play a primary role in the stratosphere-ionosphere coupling during the SSW events (e.g., Pedatella and Forbes, 2010; Goncharenko et al., 2012).

### 7 Conclusions

The presented analysis of the regional (Iberian Peninsula) atmospheric, ionospheric and geomagnetic parameters during a 2-year time interval (from July 2012 to June 2014) showed the role of the stratosphere-ionosphere coupling in the mutual variations of those parameters. While variations of the ionospheric total electron content (*iTEC*) are expected to be forced by the solar and geomagnetic activity (e.g., UV flares and geomagnetic storms), some of the *iTEC* variations, especially during the winter, characterized by a sudden stratospheric warming (SSW) event, are internally forced by the stratosphere-ionosphere coupling mechanisms.

Using the principal component analysis (PCA), we extracted a second mode of the temperature (*T*) and pressure (*gph*) variations located in the low-middle stratosphere (above ~70 hPa pressure level) that co-vary with ionospheric and geomagnetic parameters. This mode (explaining ~7% and ~3% of *T* and *gph* variations, respectively) is influenced by the global and hemispheric dynamics (QBO and polar vortex conditions) and found to be statistically significantly correlated with variations of the middle stratosphere ozone content and ionospheric *iTEC*. Similar variability of the stratospheric and ionospheric parameters was also found by the wavelet cross-coherence analysis.

To analyze the causality of the found correlations we applied the convergent cross mapping (CCM) analysis to our series, and the obtained results seem to confirm the causal character of the relations between the variations of the stratospheric (temperature and *gph*) and ionospheric (*iTEC*) parameters. During winter months, and especially during the SSW event in January 2013, the ionosphere above the analyzed region seems to be forced by the stratospheric conditions. These results are in line with a number of previous studies that using data for the middle and low latitudes of both hemispheres showed the ionospheric response to the SSW event in 2013. Nonetheless, to our knowledge, this is the first time that the coupling between the stratosphere and ionosphere is shown for the Iberian Peninsula region. This coupling is most prominent for the stratospheric temperature during easterly-QBO winter with the SSW event, and, probably, depends on the QBO-phase.

The analysis of the variations of the middle stratosphere temperature (at 10 hPa pressure level) available for the three-year time interval, 2012-2015, confirm the results obtained for the *T* mode 2. They also hint that it is a major SSW that emphasizes the stratosphere-ionosphere coupling but not the QBO phase, however longer time series are needed to confirm this hypothesis.

**Acknowledgments, Samples, and Data**

Anna Morozova was supported by the postdoc from the Fundação para a Ciência e a Tecnologia (FCT) scholarship SFRH/BPD/74812/2010. Tatiana Barlyaeva is supported by SWAIR funded by ARTES IAP Demonstration Projects. Juan José Blanco was supported through the project CTM2016-77325-C2-1-P funded by Ministerio de Economía y Competitividad.

CITEUC is funded by National Funds through FCT — Foundation for Science and Technology (project UID/Multi/00611/2013) and FEDER — European Regional Development Fund through COMPETE2020 — Operational Programme Competitiveness and Internationalization (project POCI-01-0145-FEDER-006922).

Sounding data were obtained through the GRA database http://www.ncdc.noaa.gov/data-access/weather-balloon/integrated-global-radiosonde-archiveand University of Wyoming, College of Engineering, Department of Atmospheric Science http://weather.uwyo.edu/upperair/sounding.html.

Zonally averaged stratospheric temperature and zonal wind data were downloaded from the MERRA websites http://gmao.gsfc.nasa.gov/research/merra/ and http://acdb-ext.gsfc.nasa.gov/Data_services/met/ann_data.html.

We acknowledge the use of the area-averaged data on the stratospheric temperature and ozone from AIRS Science Team/Joao Texeira, 2012, last updated 2013: AIRX3STD v006.NASA/GSFC, Greenbelt, MD, USA, NASA Goddard Earth Sciences Data and Information Services Center (GES DISC).Accessed at 10.5067/AQUA/AIRS/DATA301. The data were obtained through Giovanni online data system, developed and maintained by the NASA GES DISC http://giovanni.sci.gsfc.nasa.gov/giovanni.

The data on the QBO phases are from the Department of Earth Sciences of the Freie Universität Berlin http://www.geo.fu-berlin.de/met/ag/strat/produkte/qbo.

CaLMa cosmic ray flux data are available from NMDB website http://www.nmdb.eu/nest/search.php.

Geomagnetic data measured by the GAO UC are available by request (pribeiro@ci.uc.pt); the hourly values for the X, Y, and Z components from 2007 to 2014 can be also found at https://doi.pangaea.de/10.1594/PANGAEA.863008.

We acknowledge the use of the *Dst* index from the Kyoto World Data Center http://wdc.kugi.kyoto-u.ac.jp/dstae/index.html.

The *Mg II* data are from Institute of Environmental Physics, University of Bremen http://www.iup.uni-bremen.de/gome/gomemgii.html.

The *F10.7* data are from the OMNIweb data base https://omniweb.gsfc.nasa.gov/form/dx1.html

We also wish to thank the Ebro Observatory and Dr. Germán Solé for the provision of ionosonde data.

We acknowledge the mission scientists and principal investigators who provided the data used in this research.

We acknowledge Dr. H. Ye and his colleagues who provide implementation of the CCM analysis as an R package available at https://cran.r-project.org/package=rEDM.

**Figures captions**

**Figure 1.** Top: *Smoothed* (thin line) and *long-term* (thick line) series of *iTEC* (a), *COI H* and *Dst* (b), *Mg II* and *F1.7* (c), *O₃ 10* (d). Bottom: Altitudinal profiles (colors) of the *long-term* series of *T* (e) and *gph* (f) between 150 and 10 hPa with the subtracted mode 1 (see sec. 3.1).

**Figure 2.** Left: Results of the CCM analysis the *iTEC* vs *Mg II* for the *eQBO* (**a**) and *wQBO* (**c**) winters. The skill of cross-map estimates, indicated by the correlation coefficient ($\rho$), varies with the library length L shown as a percentage of the analyzed time interval (362 points). Red lines are reconstructions of *Mg II* from *iTEC*, blue lines are reconstructions of *iTEC* from *Mg II*. Shaded areas show 95% significance level. Wavelet cross-coherence spectrum (**b**) for the whole time interval between *Mg II* and *iTEC*. White rectangles mark *eQBO/SSW* ("1") and *wQBO/noSSW* ("2") winters. Statistically significant values are inside the black contours. Area outside of cones of influence is shaded. Right: same as left but for *iTEC* vs *COI H*.

**Figure 3.** Variations of the *Smoothed iTEC* (dark cyan lines), *Mg II* (violet lines), *COI H* (green lines) series and the *Smoothed* series of the *T* mode 2 (black solid/dashed lines for PCA/cSVD) during (**a**) *eQBO/SSW* winter (1 December, 2012 – 28 February, 2013) and (**b**) *wQBO/noSSW* winter (1 December, 2013 – 28 February, 2014). Black solid line rectangles on **a** show periods of coupled geomagnetic and ionospheric variations, whereas black dashed line rectangle shows pre-SSW and SSW period. Please note reversed Y-axes for for *COI H* on **a** and **b** and for PC2s on **b**.

**Figure 4.** Reconstructed variations of the *Smoothed T* (930-10 hPa - **a**, and 150-10 hPa - **b**) and *gph* (930-10 hPa - **c**, and 150-10 hPa - **d**) series related to the mode 2 (colors) together with corresponding PCs of the *Smoothed* series: **a** and **c**, lines with filed (PCA) and open (cSVD) symbols. Also shown: *Smoothed COI H* (**b**, green line), O₃ 10 (**c**, light grey line) and *iTEC* series (**d**, dark cyan line). On **b** and **d** the vertical black lines mark SSW event and grey lines separate *eQBO/SSW* and *wQBO/noSSW* epochs.

**Figure 5.** Left: Results of the CCM analysis of the *iTEC* vs PCs of the *noAC T* mode 2 (PCA- solid lined with solid dots, and cSVD – dashed lines with open dots) for the *eQBO/SSW* (**a**) and *wQBO/noSSW* (**c**) winters. The skill of cross-map estimates, indicated by the correlation coefficient ($\rho$), varies with the library length L shown as a percentage of the analyzed time interval (362 points). Blue lines are reconstructions of *iTEC* from *T* PCs, red lines are reconstructions of *T* PCs from *iTEC*. Shaded areas show 95% significance level (solid colors for PCA and colors with white stripes for cSVD). Wavelet cross-coherence spectrum (**b**) for the whole time interval between *T* PCA PC2 and *iTEC*. White rectangles mark *eQBO/SSW* ("1") and *wQBO/noSSW* ("2") winters. Statistically significant values are inside the black contours. Area outside of cones of influence is shaded. Right: same as left but for *gph* PCs.

**Figure 6.** (**a**) Variations of the *Smoothed T₁₀* and *T₅₀* series (grey solid and dashed lines) from 1 July, 2012 to 30 June, 2015 and *T* mode 2 (black solid/dashed lines for PCA/cSVD) from 1 July, 2012 to 30 June, 2014. (**b**) Variations of the *Smoothed iTEC* from 1 July, 2012 to 30 June, 2015.

**Figure 7.** Results of the CCM analysis of the *iTEC* vs *T₁₀* for the *eQBO/SSW* (**a**), *wQBO/noSSW* (**b**) and *eQBO/noSSW* (**c**) winters. The skill of cross-map estimates, indicated by the correlation coefficient ($\rho$), varies with the library length L shown as a percentage of the analyzed time interval (362 points). Blue lines are reconstructions of *iTEC* from *T₁₀*, red lines are reconstructions of *T₁₀* from *iTEC*. Shaded areas show 95% significance level. Please note that for the *eQBO/noSSW* winter (**c**) the $\rho(iTEC|M_{T10})$ values are negative and not shown.

**Table 1.** Correlation coefficients between *iTEC* and series of space weather parameters for the *noAC* and *Smoothed* series for data sets of different length: whole series and three winter seasons (from November to April). Only correlation coefficients $|r| \geq 0.2$ are shown. Correlation coefficients without *p values* (shown in parentheses) are statistically insignificant at 90% level.

| | whole series (July 2012 – June 2015) | *eQBO/SSW* winter (November 2012 – April 2013) | *wQBO/noSSW* winter (November 2013 – April 2014) | *eQBO/noSSW* winter (November 2014 – April 2015) |
|---|---|---|---|---|
| **(A) *noAC*** | | | | |
| *iTEC\** vs *Mg II* | 0.43 (<0.01) | 0.59 (<0.01) | 0.36 (0.08) | 0.22 |
| *iTEC\** vs *F10.7* | 0.34 (<0.01) | 0.56 (<0.01) | | |
| *iTEC vs COI H* | | -0.26 (<0.01) | | |
| *iTEC vs Dst* | | -0.22 (0.07) | | |
| **(B) *Smoothed*** | | | | |
| *iTEC vs Mg II* | 0.52 (<0.01) | 0.66 (<0.01) | 0.41 (0.05) | 0.26 |
| *iTEC vs F10.7* | 0.41 (<0.01) | 0.63 (<0.01) | | |
| *iTEC vs COI H* | | -0.34 (0.03) | | 0.21 |
| *iTEC vs Dst* | | -0.24 | | |

**Table 2.** PCA results for the *T* and *gph* series (mode 2 only).

| parameter | series | PCA | | cSVD | |
|---|---|---|---|---|---|
| | | number | variance fraction | number | variance fraction, % |
| *T* | *noAC* | 4 | 6% | 4 | 4% |
| | *Smoothed* | 2 | 15% | 4 | 4% |
| *gph* | *noAC* | 4 | 2.5% | 4 | 4% |
| | *Smoothed* | 5 | 1% | 4 | 4% |

**Table 3.** Same as Table 1 but for *iTEC* and series of atmospheric parameters: mode 2 (from PCA and cSVD) of the *T* and *gph* series and the $T_{10\,hPa}$ and $T_{50\,hPa}$ series. Winter seasons only.

| | *eQBO/SSW* winter (November 2012 – April 2013) | *wQBO/noSSW* winter (November 2013 – April 2014) | *eQBO/noSSW* winter (November 2014 – April 2015) |
|---|---|---|---|
| *iTEC vs $T_{PCA}$* | 0.33 (0.04) | -0.38 | *no data* |
| *iTEC vs $T_{cSVD}$* | 0.33 (0.01) | -0.32 | *no data* |
| *iTEC vs $gph_{PCA}$* | 0.33 (0.01) | -0.39 (0.03) | *no data* |
| *iTEC vs $gph_{cSVD}$* | 0.37 (0.01) | -0.3 | *no data* |
| *iTEC vs $T_{10\,hPa}$* | 0.27 (0.1) | -0.52 (0.04) | -0.24 (0.06) |
| *iTEC vs $T_{50\,hPa}$* | 0.32 (0.03) | -0.27 | -0.24 (0.1) |
| *iTEC vs $T_{PCA}$* | 0.37 (0.05) | -0.42 | *no data* |
| *iTEC vs $T_{cSVD}$* | 0.41 (0.01) | -0.36 | *no data* |
| *iTEC vs $gph_{PCA}$* | 0.37 (0.02) | -0.45 (0.04) | *no data* |
| *iTEC vs $gph_{cSVD}$* | 0.29 (0.1) | -0.57 (0.04) | *no data* |
| *iTEC vs $T_{10\,hPa}$* | 0.35 (0.04) | -0.31 | -0.27 (0.09) |
| *iTEC vs $T_{50\,hPa}$* | 0.42 (0.01) | -0.34 | -0.28 |

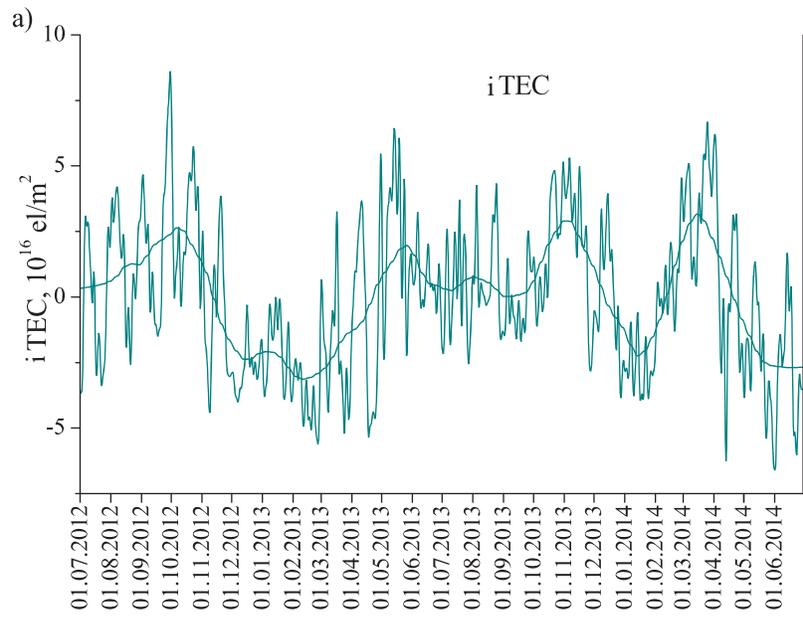

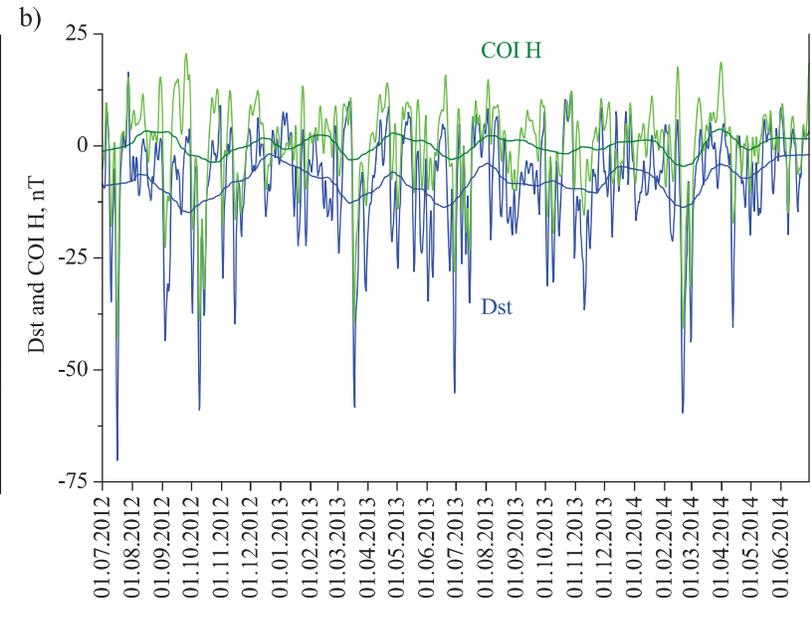

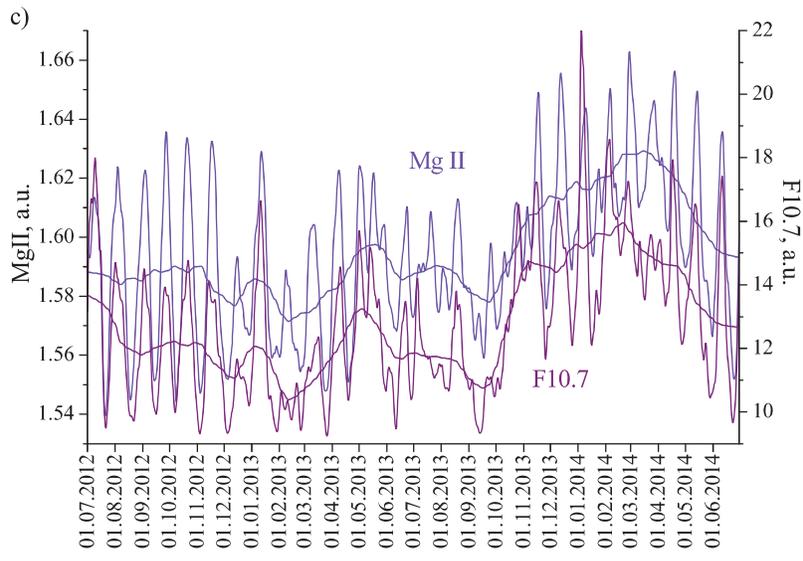

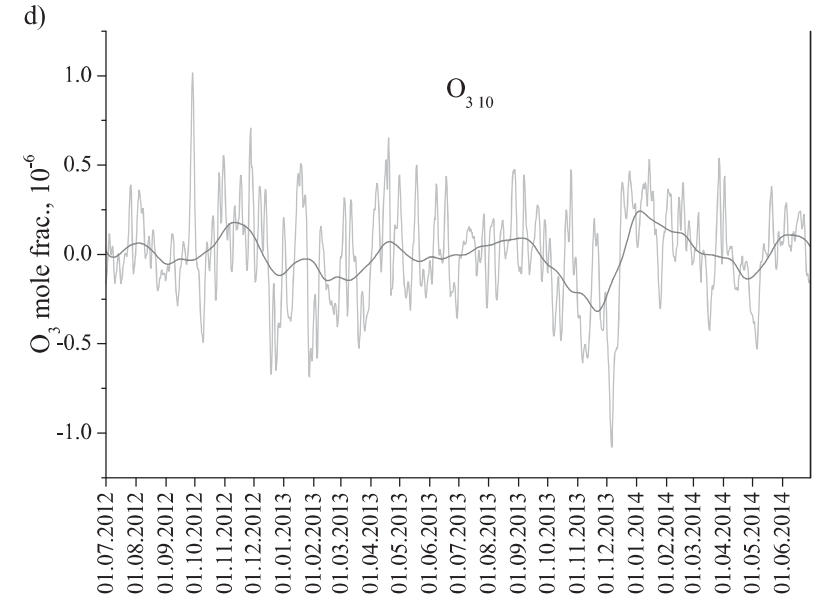

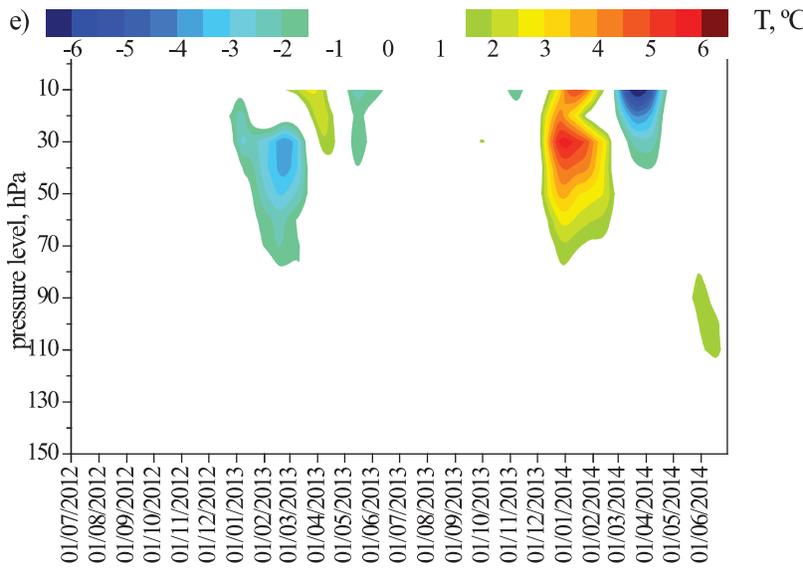

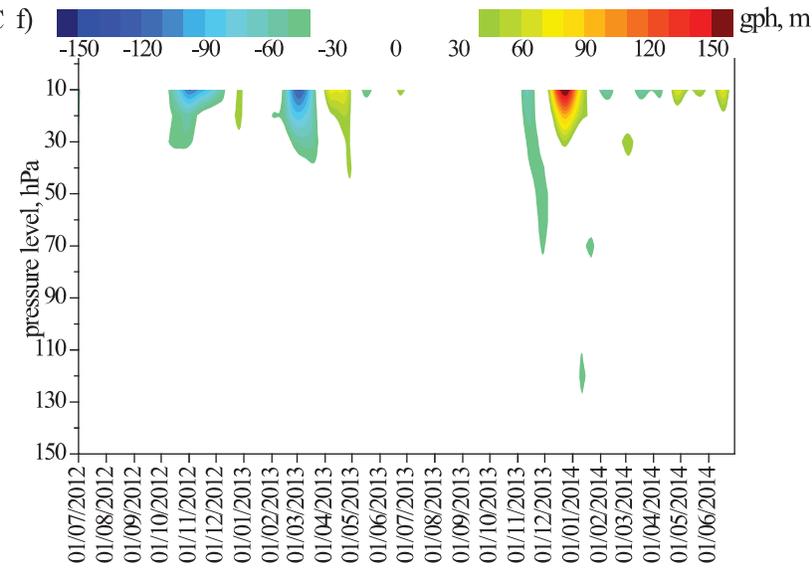

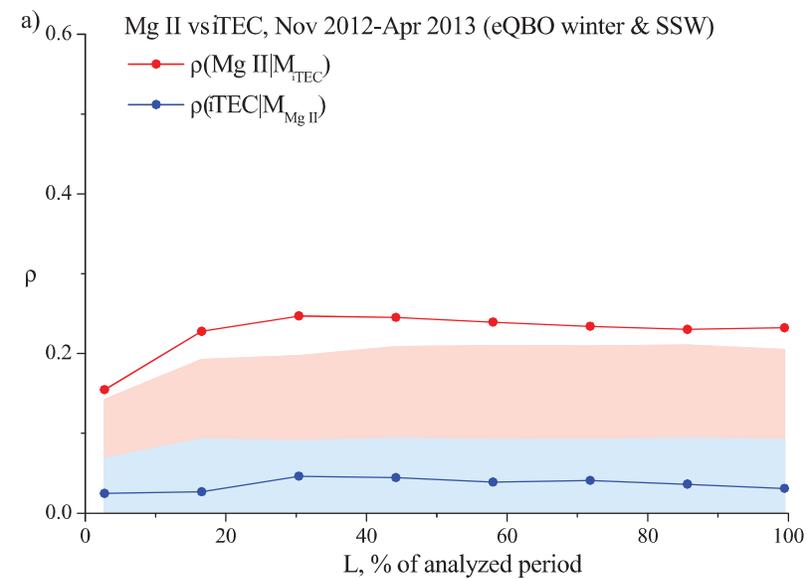

a) Mg II vsiTEC, Nov 2012-Apr 2013 (eQBO winter & SSW)

$\rho(Mg\ II|M_{iTEC})$

$\rho(iTEC|M_{Mg\ II})$

L, % of analyzed period

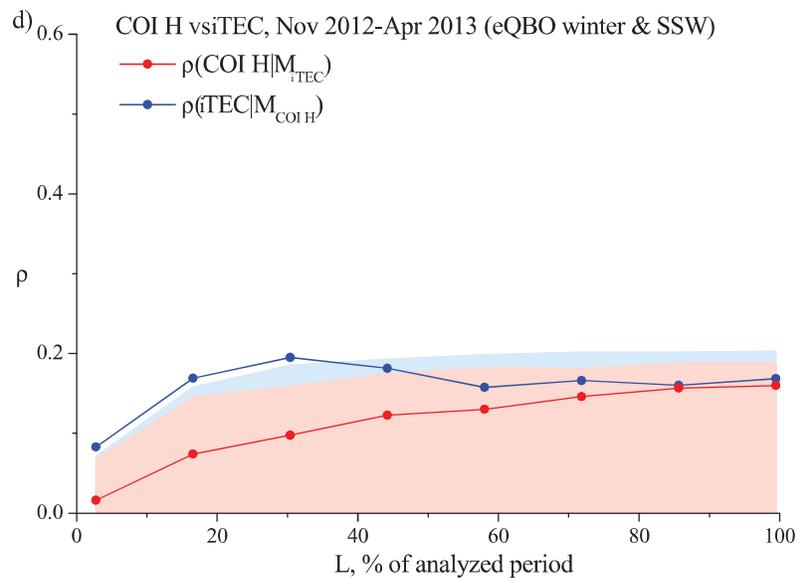

d) COI H vsiTEC, Nov 2012-Apr 2013 (eQBO winter & SSW)

$\rho(COI\ H|M_{iTEC})$

$\rho(iTEC|M_{COI\ H})$

L, % of analyzed period

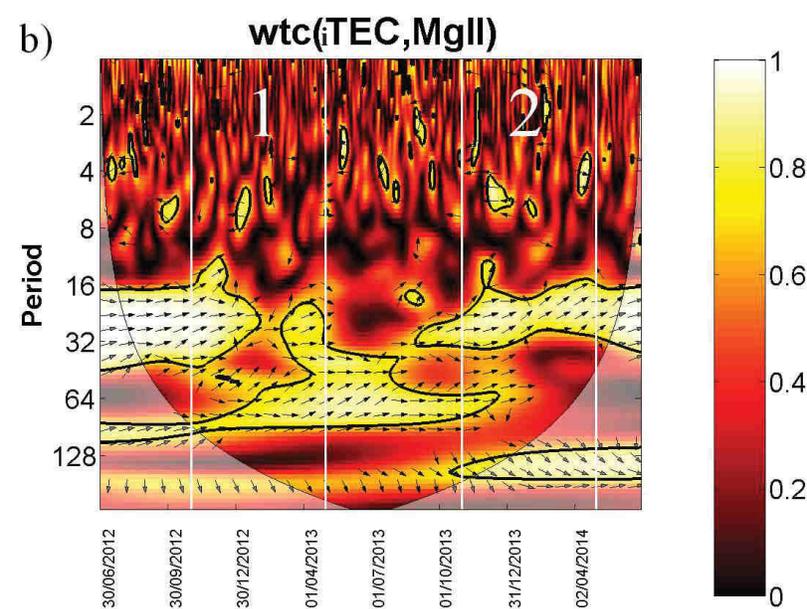

b) **wtc(iTEC,MgII)**

Period

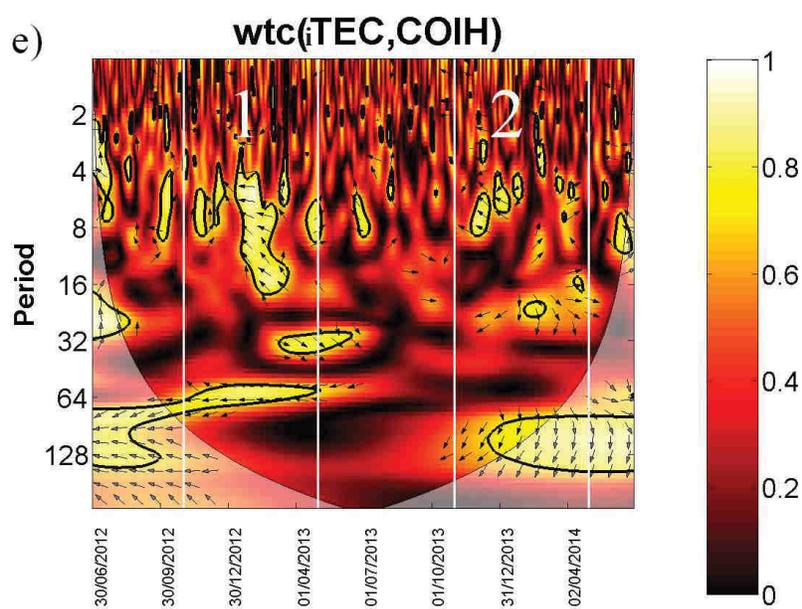

e) **wtc(iTEC,COIH)**

Period

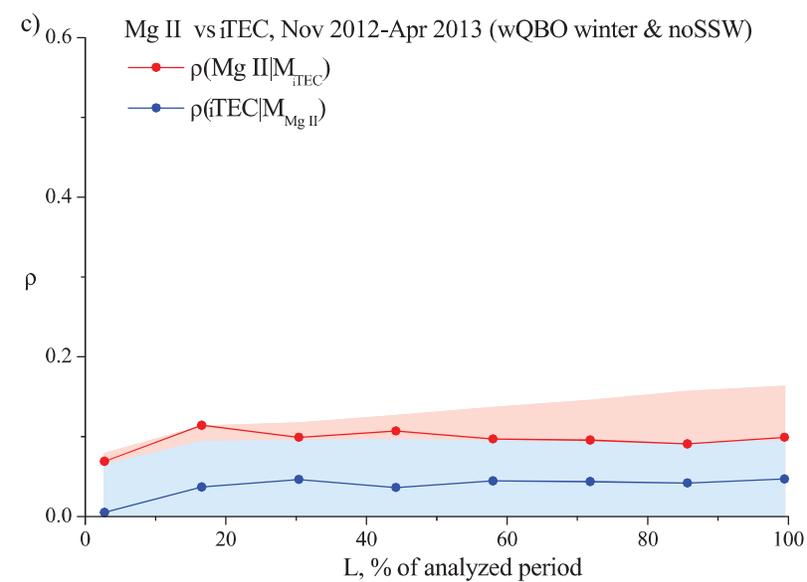

c) Mg II vs iTEC, Nov 2012-Apr 2013 (wQBO winter & noSSW)

$\rho(Mg\ II|M_{iTEC})$

$\rho(iTEC|M_{Mg\ II})$

L, % of analyzed period

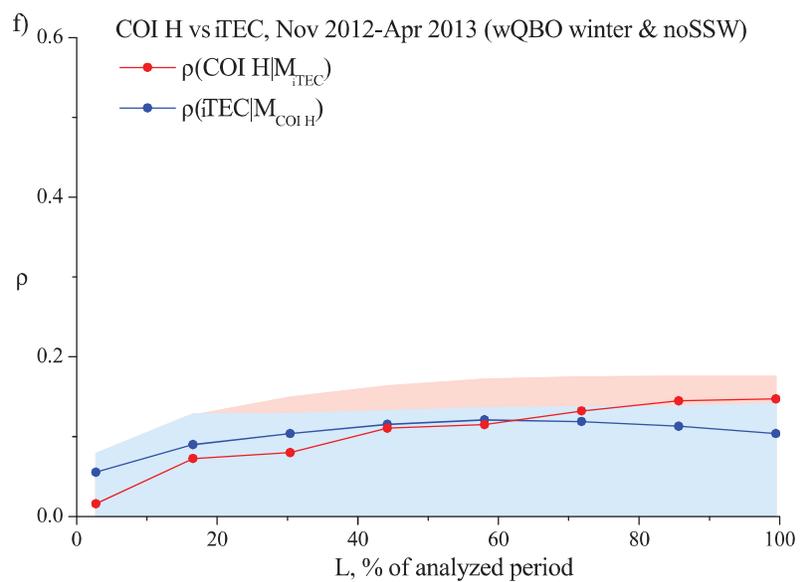

f) COI H vs iTEC, Nov 2012-Apr 2013 (wQBO winter & noSSW)

$\rho(COI\ H|M_{iTEC})$

$\rho(iTEC|M_{COI\ H})$

L, % of analyzed period

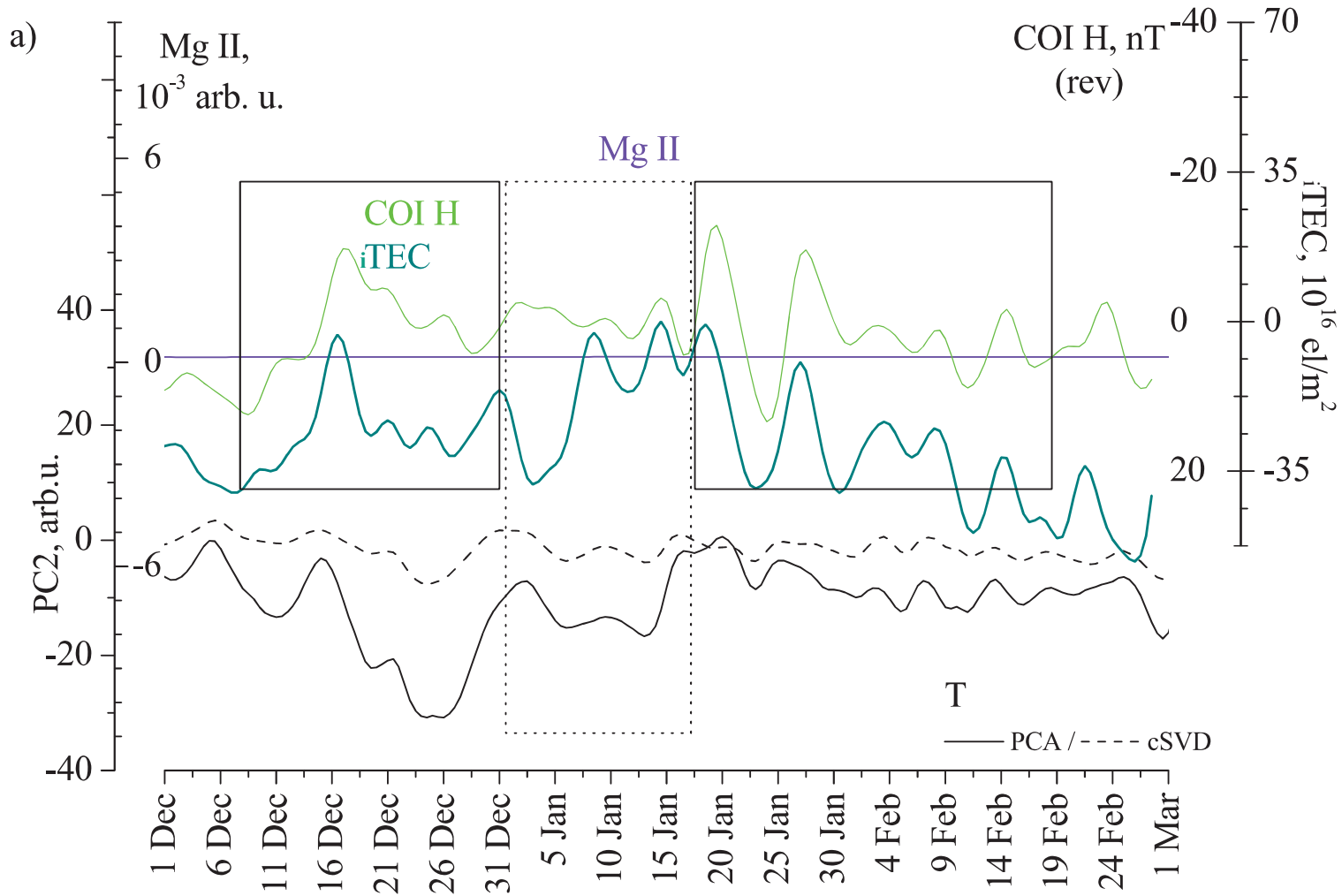

a)

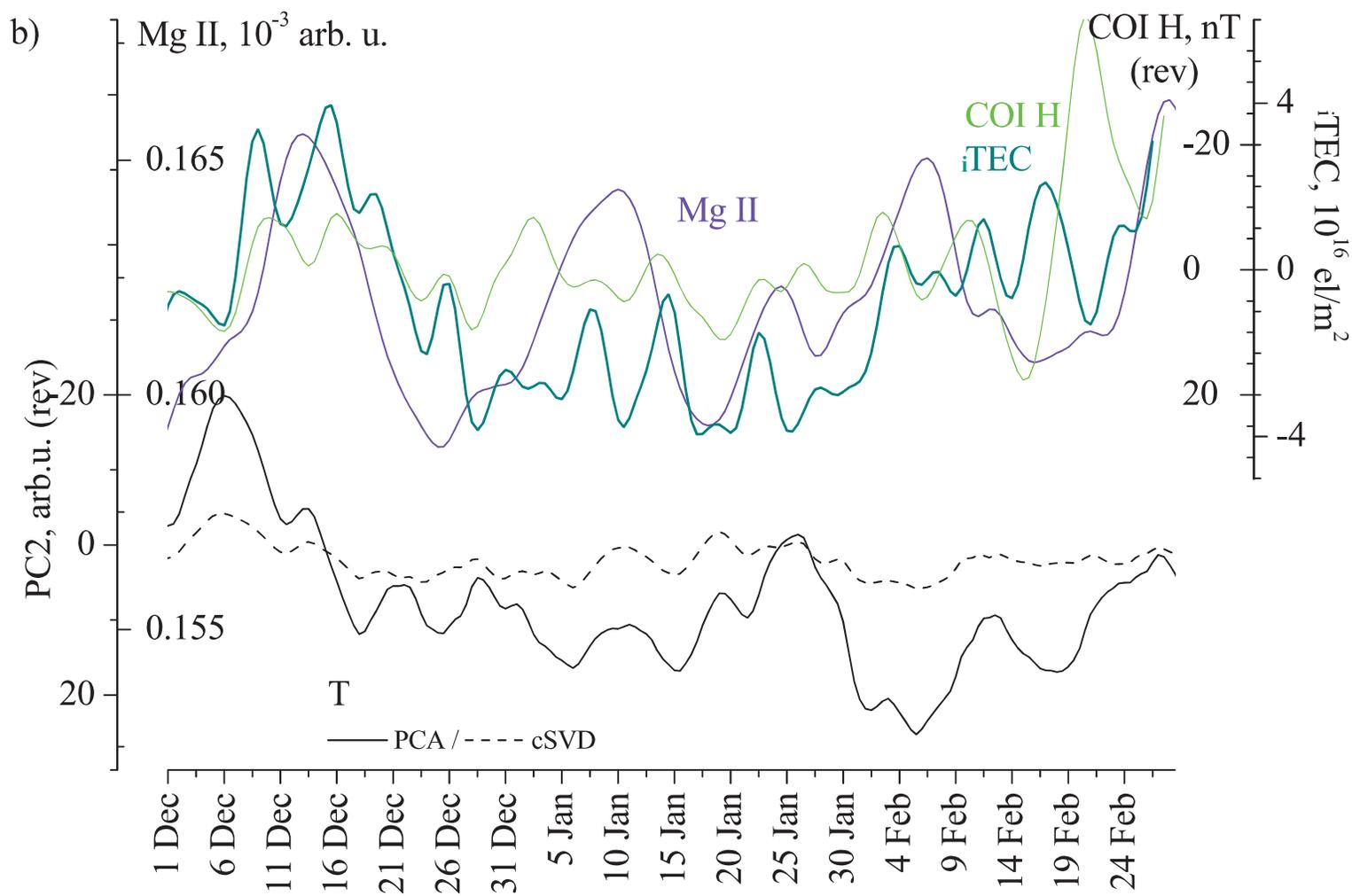

b)

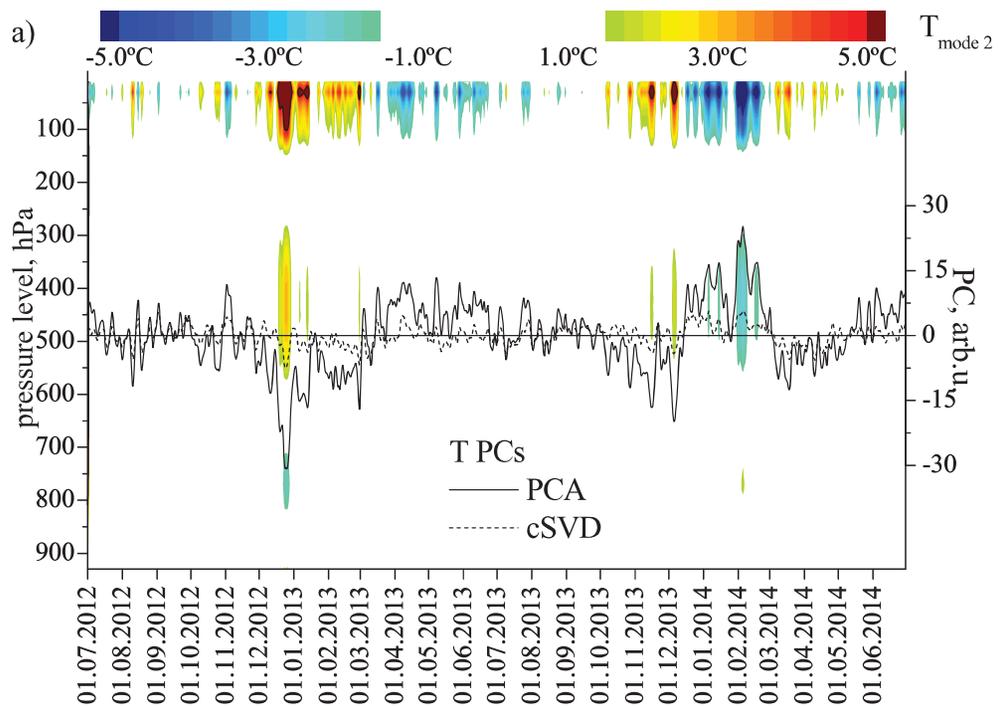

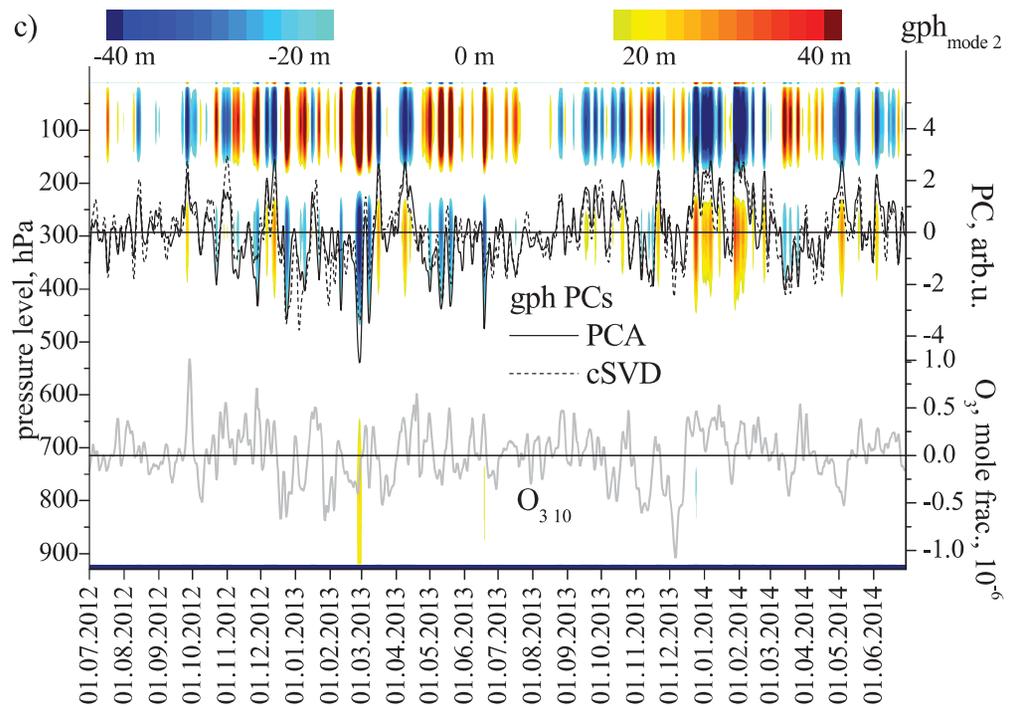

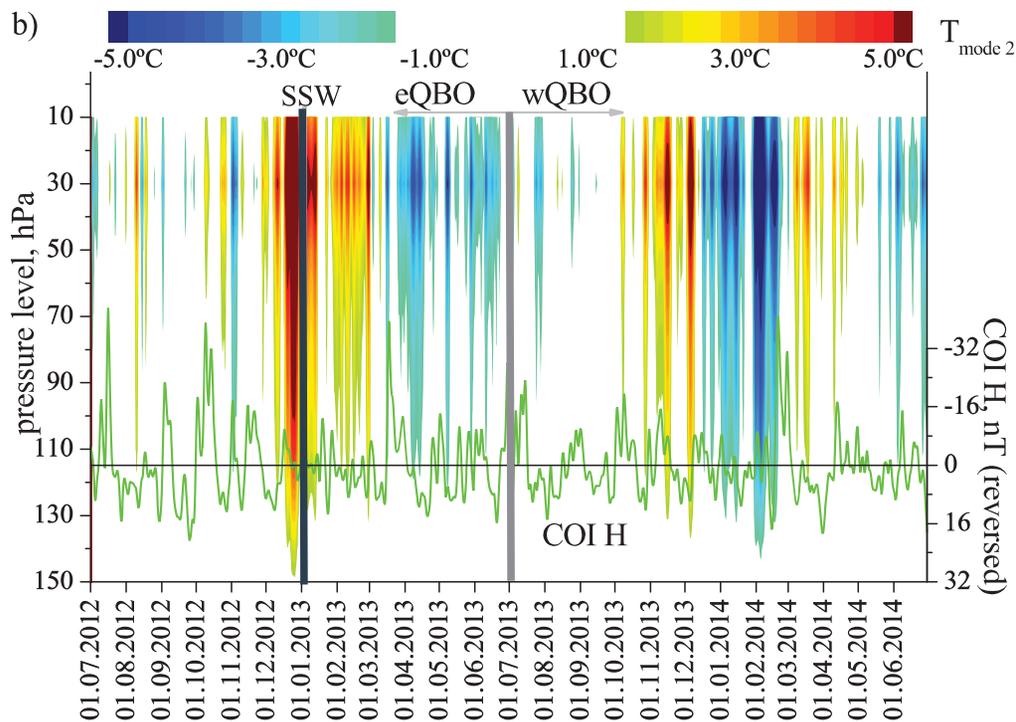

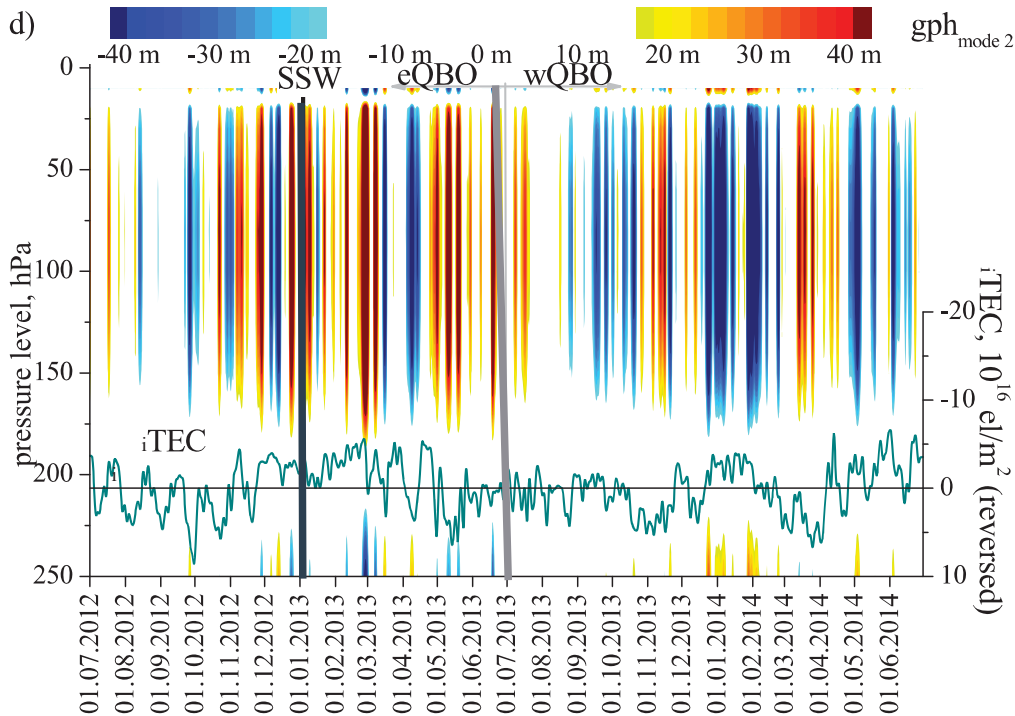

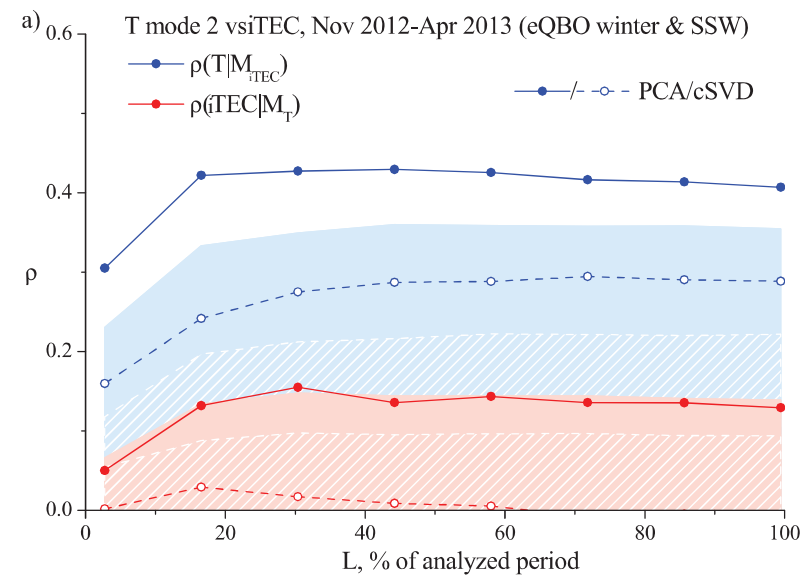

a)

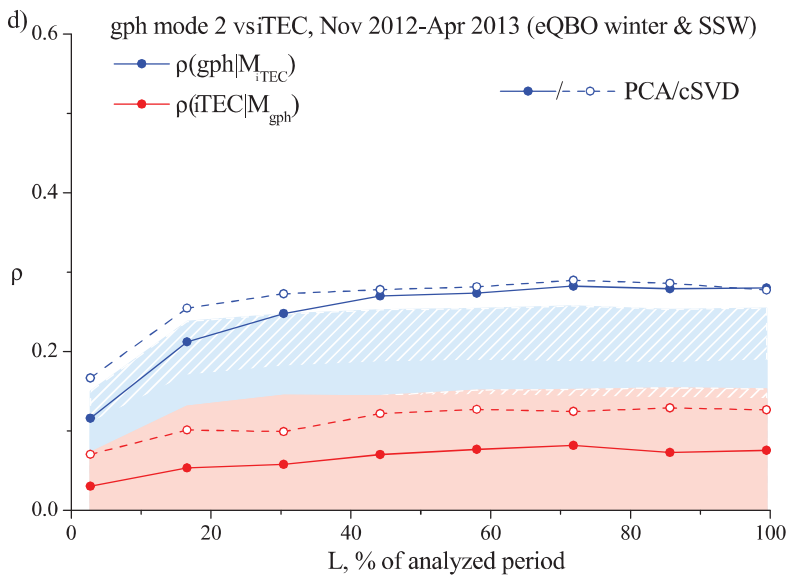

d)

i

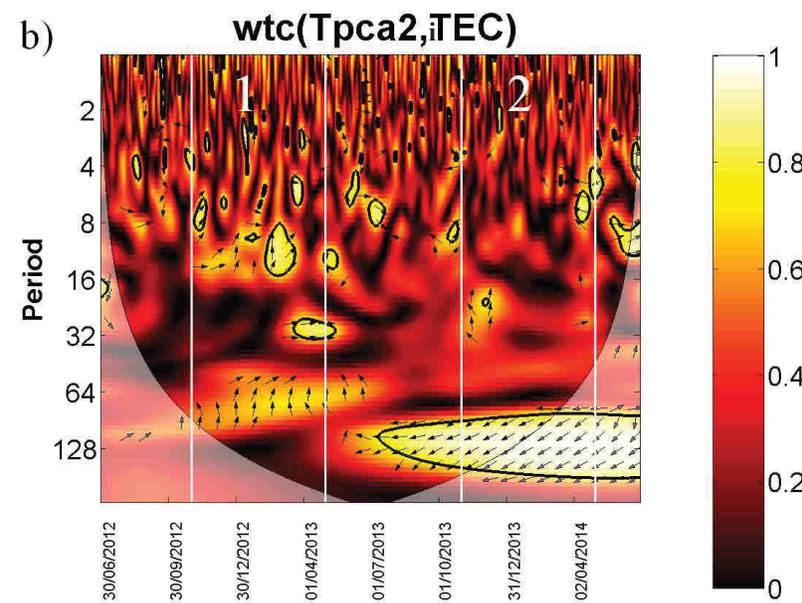

b) **wtc(Tpca2,iTEC)**

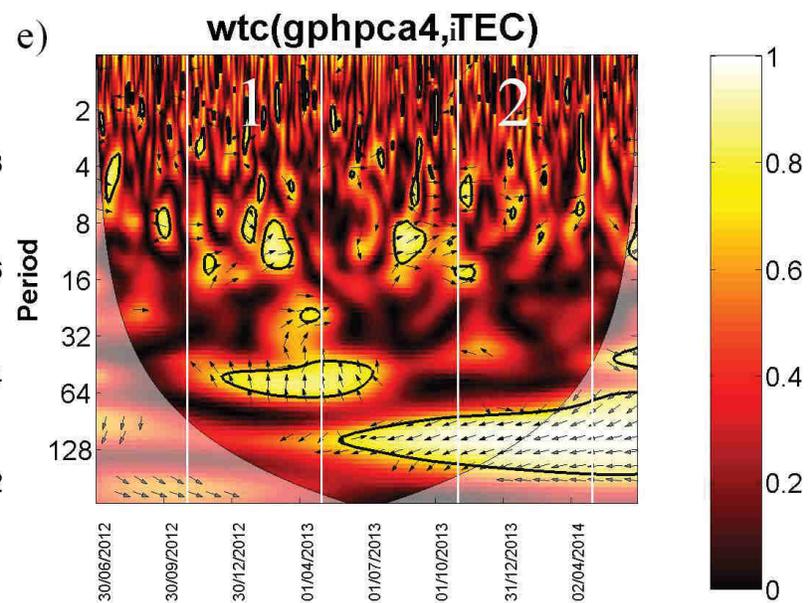

e) **wtc(gphpca4,iTEC)**

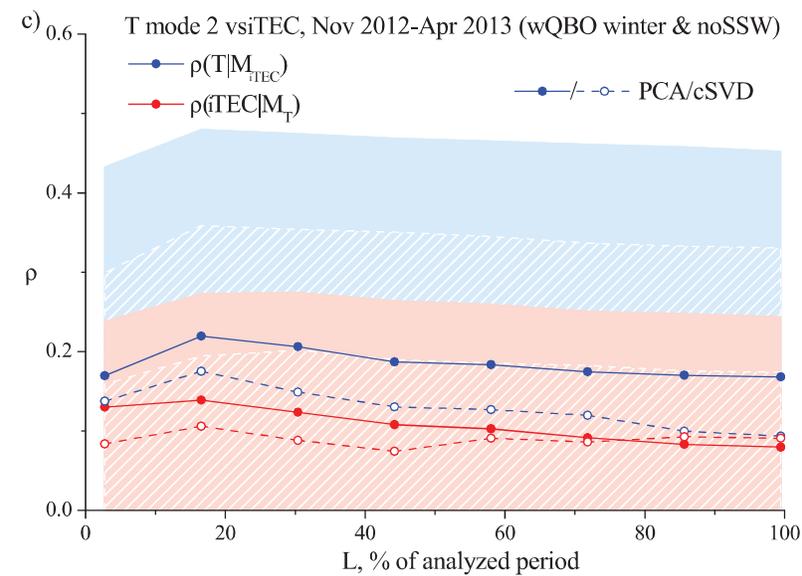

c)

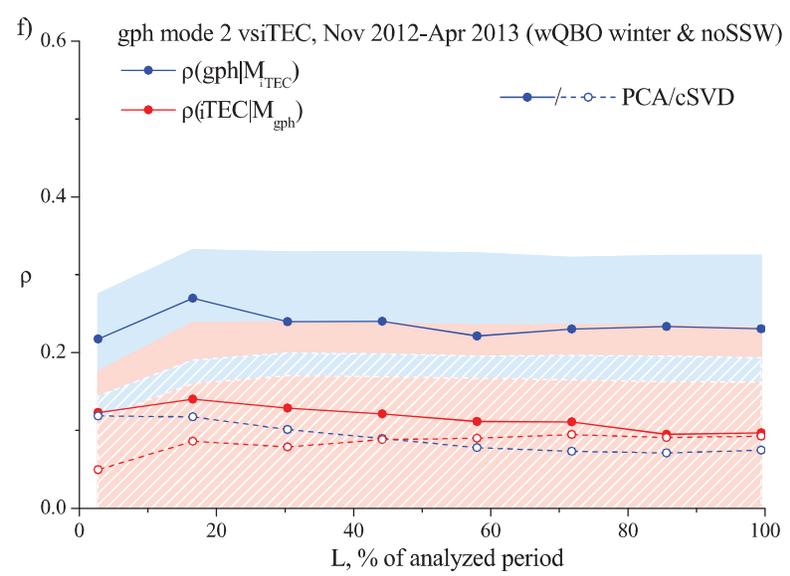

f)

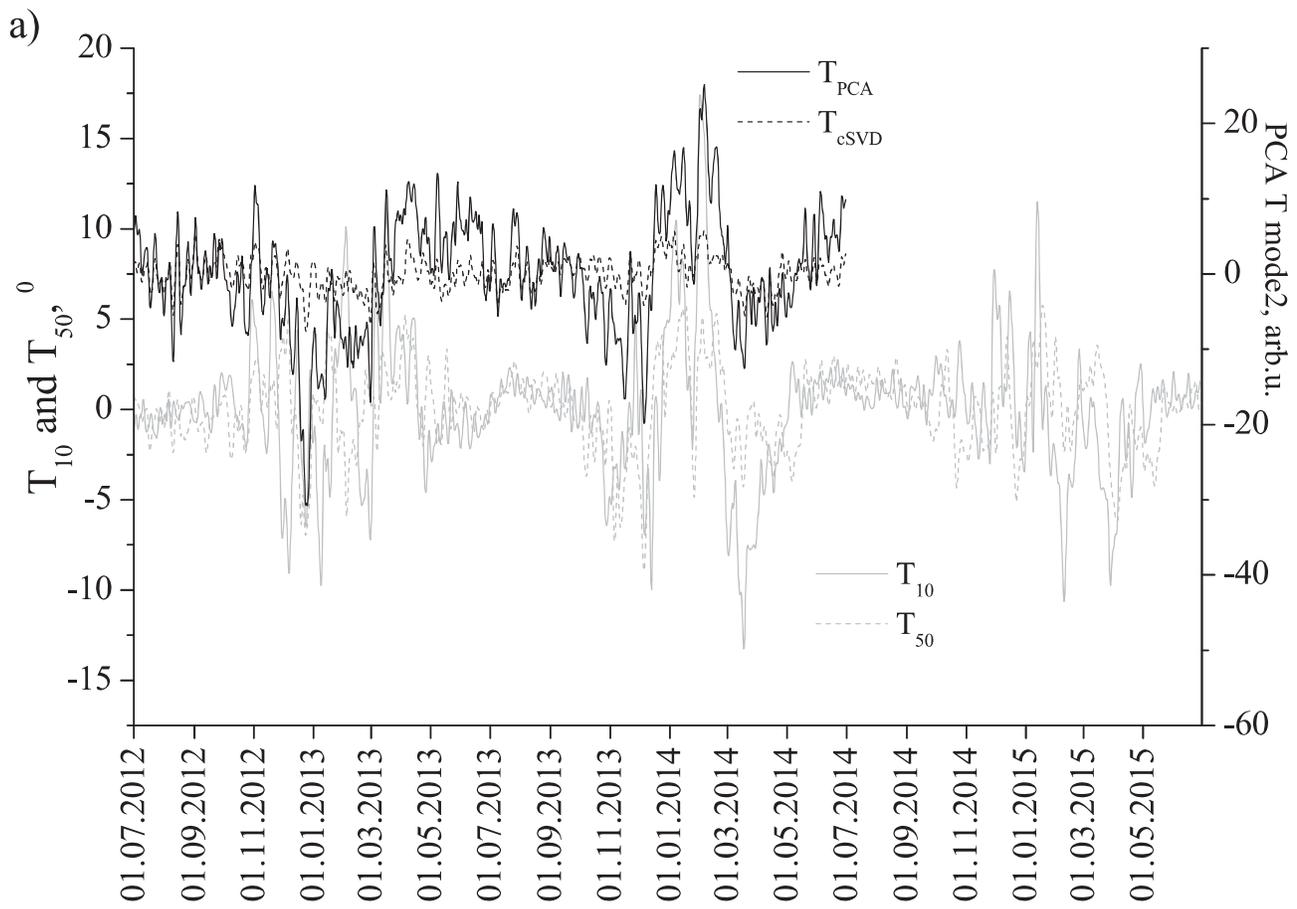

a)

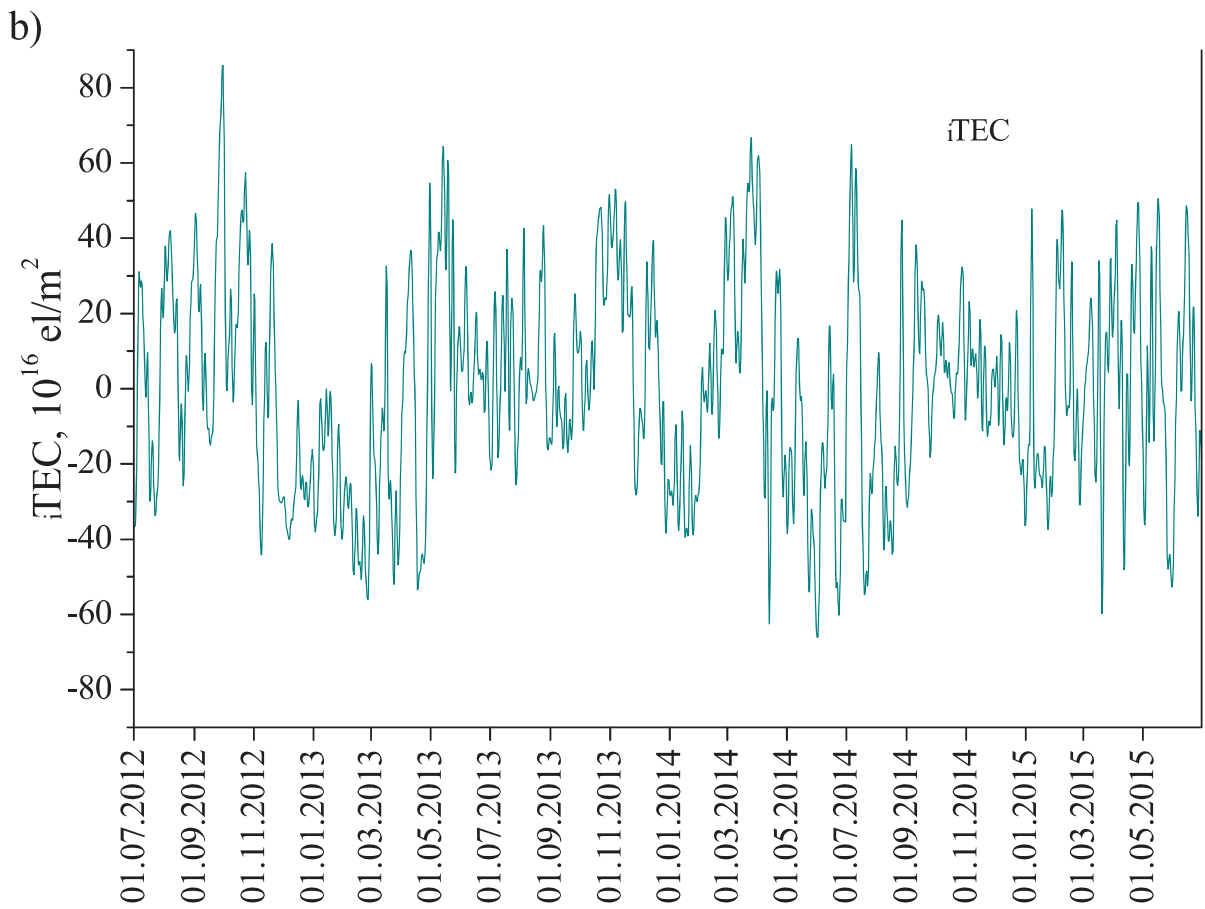

b)

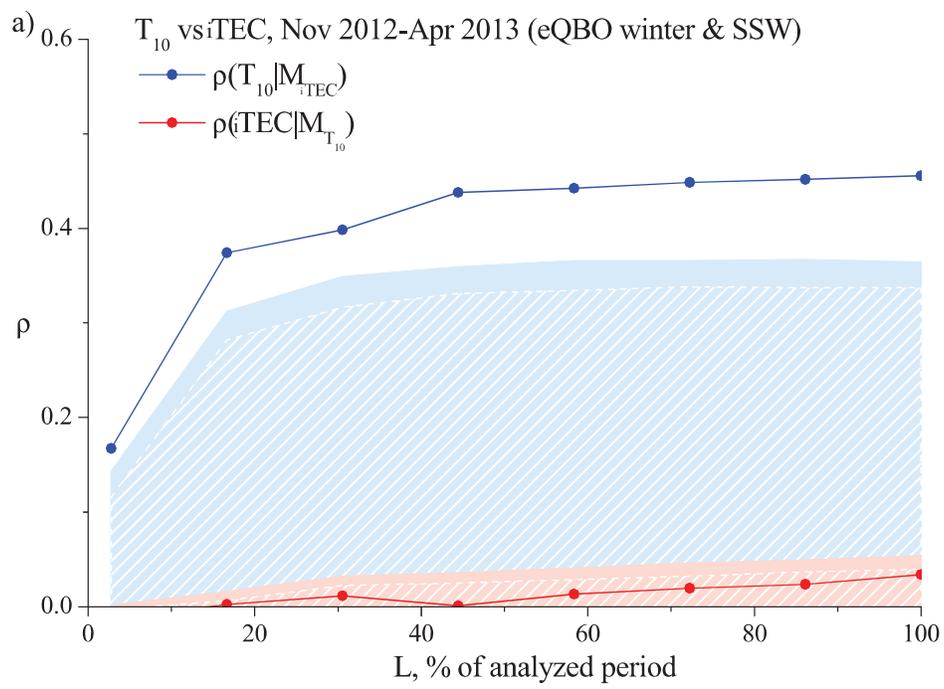

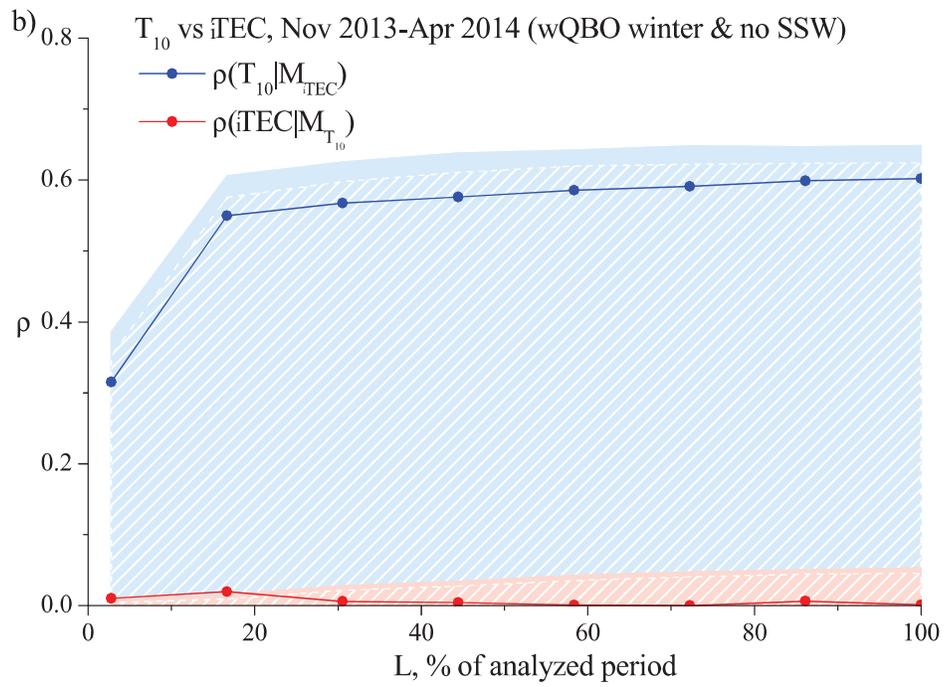

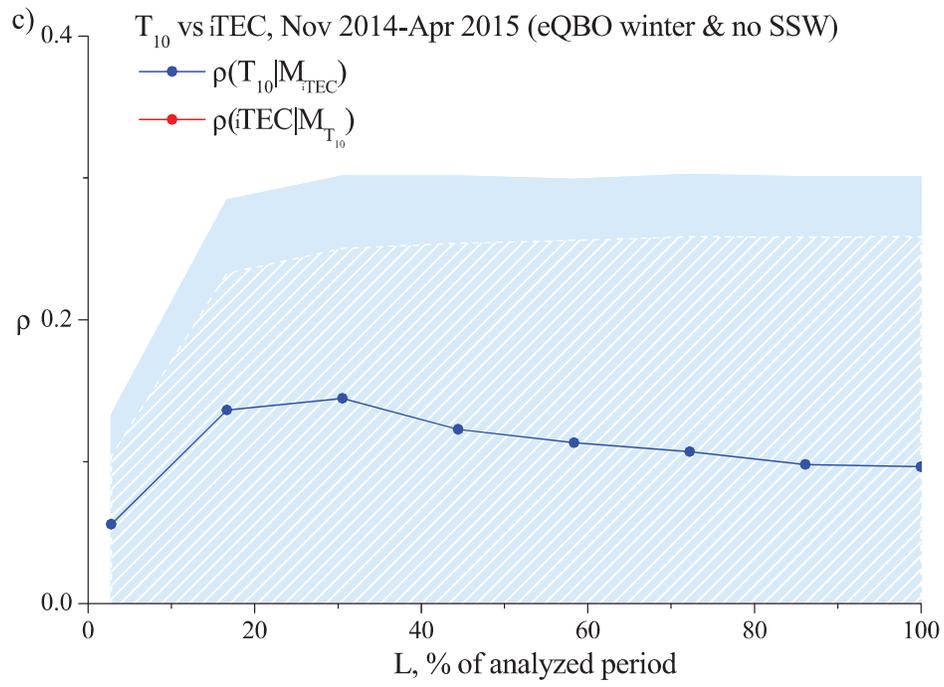